\documentclass[preprint,12pt]{elsarticle}

\usepackage{graphicx}
\usepackage{cases}
\usepackage{amssymb}
\usepackage{amsthm}
\newtheorem{proposition}{Proposition}

\usepackage{color}

\journal{Journal of Computational Physics}

\begin{document}

\begin{frontmatter}

\title{Biot-JKD model: simulation of 1D transient \\ poroelastic waves with fractional derivatives}

\author[LMA]{Emilie Blanc}
\ead{eblanc@lma.cnrs-mrs.fr}
\author[M2P2]{Guillaume Chiavassa  \corref{cor1}}
\ead{guillaume.chiavassa@centrale-marseille.fr}
\author[LMA]{Bruno Lombard}
\ead{lombard@lma.cnrs-mrs.fr}
\cortext[cor1]{Corresponding author. Tel.: +33 491 05 46 69.}
\address[LMA]{Laboratoire de M\'{e}canique et d'Acoustique, UPR 7051 CNRS, 31 chemin Joseph Aiguier, 13402 Marseille, France}
\address[M2P2]{Centrale Marseille and M2P2, UMR 7340 - CNRS,
Technop\^ole de Chateau-Gombert, 38 rue Fr\'ed\'eric Joliot-Curie, 13451 Marseille, France}

\begin{abstract}
A time-domain numerical modeling of Biot poroelastic waves is presented. The viscous dissipation occurring in the pores is described using the dynamic permeability model developed by Johnson-Koplik-Dashen (JKD). Some of the coefficients in the Biot-JKD model are proportional to the square root of the frequency: in the time-domain, these coefficients introduce order $1/2$ shifted fractional derivatives involving a convolution product. Based on a diffusive representation, the convolution kernel is replaced by a finite number of memory variables that satisfy local-in-time ordinary differential equations. Thanks to the dispersion relation, the coefficients in the diffusive representation are obtained by performing an optimization procedure in the frequency range of interest. A splitting strategy is then applied numerically: the propagative part of Biot-JKD equations is discretized using a fourth-order ADER scheme on a Cartesian grid, whereas the diffusive part is solved exactly. Comparisons with analytical solutions show the efficiency and the accuracy of this approach.
\end{abstract}


\end{frontmatter}



\section{Introduction}\label{SecIntro}

Porous media consist of a solid matrix within which fluids can circulate freely. The propagation of waves in these media has many crucial implications in applied mechanics, in situations where materials such as industrial foams, spongious bones \cite{SEBAA06} and petroleum rocks \cite{BOURBIE87} have to be characterized, for example. The poroelastic model originally developed by Biot in 1956 \cite{BIOT56-A} includes two classical waves (one "fast" compressional wave and one shear wave), in addition to a second "slow" compressional wave, which is highly dependent on the saturating fluid. This slow wave was observed experimentally in 1981 \cite{PLONA80}, thus confirming the validity of Biot's theory.

Two frequency regimes have to be distinguished when dealing with poroelastic waves. One of the main problems is how to model the dissipation of mechanical energy. In the low-frequency range (LF) \cite{BIOT56-A}, the viscous boundary layer that develops in the fluid is large in comparison with the diameter of the pores, and the viscous efforts are proportional to the relative velocity of the motion between the fluid and solid components. In the high-frequency range (HF), modeling the dissipation is a more delicate task: Biot first presented an expression for particular pore geometries \cite{BIOT56-B}. In 1987, Johnson-Koplik-Dashen (JKD) \cite{JKD87} published a general expression for the dissipation in the case of random pores. The viscous efforts depend in this model on the square root of the frequency of the perturbation. When writing the evolution equations in the time domain, time fractional derivatives are introduced, which involves convolution products with singular kernels \cite{LUBICH86}. Analytical solutions have been derived in simple academic geometries and homogeneous media \cite{FELLAH04}.

Many numerical methods have been developed in the LF regime: see \cite{CARCIONE10} and the introduction to \cite{CHIAVASSA11} for general reviews. In the HF regime, the fractional derivatives greatly complicate the numerical modeling of the Biot-JKD equations. The past values of the solution are indeed required in order to evaluate these convolution products, which means that the time evolution of the solution must be stored. This of course greatly increases the memory requirements and makes large-scale simulations impossible. To our knowledge, only two approaches to this problem have been proposed so far in the literature. The first approach consisted in discretizing the convolution products \cite{MASSON10}, and the second one was based on the use of a diffusive representation of the fractional derivative \cite{HANYGA05,TORRES96}. In the latter approach, the convolution product is replaced by a continuum of diffusive variables - or memory variables - satisfying local differential equations \cite{MATIGNON10}. This continuum is then discretized using appropriate quadrature formulas, resulting in the Biot-DA (diffusive approximation) model. 

However, the diffusive approximation proposed in \cite{HANYGA05} has three major drawbacks. First, the quadrature formulas make the convergence towards the original fractional operator very slow. Secondly, in the case of small frequencies, the Biot-DA model does not converge towards the Biot-LF model. Lastly, the number of memory variables required is not specified. The aim of the present study is therefore to develop a new diffusive approximation method in which these drawbacks do not arise. Since it is proposed here to focus on the discretization of the fractional derivatives, we will deal only with the 1-D  equations of evolution in homogeneous media, so that the shear wave will not be considered. However, the strategy proposed here can be extended quite straightforwardly  to 2D and 3D geometries, as discussed below.

This paper is organized as follows. The original Biot-JKD model is briefly outlined in section \ref{SecPhys} and the principles underlying the diffusive representation of fractional derivatives are described. The decrease of energy and the dispersion analysis are addressed. In section \ref{SecDA}, the method used to discretize the diffusive model is presented: the diffusive approximation thus obtained is easily treatable by computers. Following a similar approach than in viscoelasticity \cite{GROBY11}, the coefficients of the model are determined using an optimization procedure in the frequency range of interest, giving an optimum number of additional computational arrays. The numerical modeling is addressed in section \ref{SecNum}, where the equations of evolution are split into two parts: a propagative part, which is discretized using a fourth-order scheme for hyperbolic equations, and a diffusive part, which is solved exactly. Some numerical experiments performed with realistic values of the physical parameters are presented in section \ref{SecExp}. In section \ref{SecConclu}, a conclusion is drawn and some futures lines of research are given. 
 

\section{Physical modeling}\label{SecPhys}

\subsection{Biot model}\label{SecPhysBiot}

The Biot model describes the propagation of mechanical waves in a macroscopic porous medium consisting of a solid matrix saturated with a fluid circulating freely through the pores \cite{BIOT56-A,BOURBIE87,CARCIONE07}. It is assumed that
\begin{itemize}
\item the wavelengths are large in comparison with the diameter of the pores;
\item the amplitude of the perturbations is small;
\item the elastic and isotropic matrix is completely saturated with a single fluid phase;
\item the thermo-mechanical effects are neglected.
\end{itemize}
This model involves 10 physical parameters: the density $\rho_f$ and the dynamic viscosity $\eta$ of the fluid; the density $\rho_s$ and the shear modulus $\mu$ of the elastic skeleton; the porosity $0<\phi<1$, the tortuosity $a\geq 1$, the absolute permeability at null frequency $\kappa$, the Lam\'e coefficient $\lambda_f$ and the two Biot's coefficients $\beta$ and $m$ of the saturated matrix. The following notations are introduced
\begin{equation}
\begin{array}{l}
\displaystyle
\rho_w = \frac{a}{\phi}\,\rho_f\mbox{,}\quad \rho = \phi\,\rho_f + (1-\phi)\,\rho_s\mbox{,}\quad \chi = \rho\,\rho_w - \rho_f^2 >0,\\
[10pt]
\displaystyle
\quad\lambda_0 = \lambda_f-m\,\beta^2,\quad C=\lambda_0+2\,\mu>0.
\end{array}
\label{parametre_biot}
\end{equation}
Taking $u_s$ and $u_f$ to denote the solid and fluid displacements, the unknowns in 1D are the elastic velocity $v_s=\frac{\partial\,u_s}{\partial\,t}$, the filtration velocity $w=\frac{\partial\,{\cal W}}{\partial\,t}=\phi\,\frac{\partial }{\partial\,t}\,(u_f-u_s)$, the elastic stress $\sigma$, and the acoustic pressure $p$. The constitutive laws are
\begin{subnumcases}{\label{biot_comportement}}
\displaystyle \sigma = (\lambda_f+2\,\mu)\,\varepsilon -m\,\beta\,\xi,\\
[10pt]
\displaystyle p = m\,(-\beta\,\varepsilon + \xi),
\end{subnumcases}
where $\varepsilon = \frac{\partial u_s}{\partial x}$ is the strain and $\xi = -\frac{\partial \mathcal{W}}{\partial x}$ is the rate of fluid change. On the other hand, the conservation of momentum yields
\begin{subnumcases}{\label{biot_dynamique}}
\displaystyle \rho\,\frac{\partial v_s}{\partial t} + \rho_f\,\frac{\partial w}{\partial t} = \frac{\partial \sigma}{\partial x}+f_b,\\
[10pt]
\displaystyle \rho_s\,\frac{\partial v_s}{\partial t} + \rho_w\,\frac{\partial w}{\partial t} + \frac{\eta}{\kappa}\,F*w = -\frac{\partial p}{\partial x}+f_f,\label{darcy}
\end{subnumcases}
where $*$ is the convolution product in time; $f_b$ and $f_f$ are the body force for an unit volume of the bulk material and the pore fluid, respectively. The equation (\ref{darcy}) is a generalized Darcy law. The quantity $F*w$  denotes the viscous dissipation induced by the relative motion between the fluid and the elastic skeleton.


\subsection{High frequency dissipation: the JKD model}\label{SecPhysJKD}

The frontier between the low-frequency range (LF) and the high-frequency range (HF) is reached when the viscous efforts and the inertial effects are similar. The transition frequency is given by \cite{BIOT56-A,BOURBIE87} 
\begin{equation}
f_c = \frac{\eta\,\phi}{2\,\pi\,a\,\kappa\,\rho_f}=\frac{\textstyle \omega_c}{\textstyle 2\,\pi}.
\label{Fc}
\end{equation}
In LF, the flow in the pores is of the Poiseuille type, and dissipation efforts in (\ref{darcy}) are given by
\begin{equation}
F(t)= \delta(t) \Longleftrightarrow F(t)*w(x,\,t)=w(x,\,t),
\end{equation}
where $\delta$ is the Dirac distribution. In HF, the width of the viscous boundary-layer is small in comparison with the size of the pores, and modeling the dissipation process is a more complex task. Here we adopt the widely-used model proposed by Johnson-Koplik-Dashen (JKD) in 1987, which is valid for random networks of pores with constant radii \cite{JKD87}. The only additional parameter is the viscous characteristic length $\Lambda$. We take
\begin{equation}
P=\frac{4\,a\,\kappa}{\phi\,\Lambda^2}\mbox{,}\qquad\Omega = \frac{\omega_c}{P} = \frac{\eta\,\phi^2\,\Lambda^2}{4\,a^2\,\kappa^2\,\rho_f},
\label{coef_hf}
\end{equation}
where $P$ is the Pride number (typically $P \approx 1/2$). Based on the Fourier transform in time, $\widehat{F}(\omega) = \int _{\mathbb R} F(t)e^{-i\omega t}\,dt$, the frequency correction given by the JKD model can be written
\begin{equation}
\begin{array}{ll}
\widehat{F}(\omega) & \displaystyle = \left( 1+i\,\omega\,\frac{4\,a^2\,\kappa^2\,\rho_f}{\eta\,\Lambda^2\,\phi^2}\right) ^{1/2},\\
[10pt]
& \displaystyle = \left( 1+i\,P\,\frac{\omega}{\omega_c}\right) ^{1/2},\\
[10pt]
& \displaystyle = \frac{1}{\sqrt{\Omega}}\,(\Omega +i\,\omega)^{1/2}.
\end{array}
\label{F_omega}
\end{equation}
This correction is the simplest function satisfying the LF and HF limits of the dynamic permeability \cite{JKD87}. Therefore, the term $F(t)*w(x,t)$ involved in (\ref{darcy}) is
\begin{equation}
\begin{array}{ll}
F(t)*w(x,t) & \displaystyle = \mathcal{F}^{-1}\left( \frac{1}{\sqrt{\Omega}}\,(\Omega + i\,\omega)^{1/2}\widehat{w}(x,\omega)\right),\\
[10pt]
& \displaystyle = \frac{1}{\sqrt{\Omega}}\,(D+\Omega)^{1/2}w(x,t).
\end{array}
\label{F_t}
\end{equation}
The operator $D^{1/2}$ is a shifted order 1/2 time fractional derivative, generalizing the usual derivative characterized by $\frac{\partial\,w}{\partial\,t} = \mathcal{F}^{-1}\left( i\,\omega\,\widehat{w}(\omega)\right)$. The notation $(D+\Omega)^{1/2}$ accounts for the shift $\Omega$ in (\ref{F_t}).


\subsection{The Biot-JKD equations of evolution}\label{SecPhysEDP}

Based on (\ref{biot_comportement}), (\ref{biot_dynamique}) and (\ref{F_t}), the Biot-JKD equations can be written
\begin{subnumcases}{\label{LCBiot}}
\displaystyle \rho\,\frac{\partial\,v_s}{\partial\,t} + \rho_f\,\frac{\partial\,w}{\partial\,t}=\frac{\partial\,\sigma}{\partial\,x}+f_b,\label{LCBiot_v}\\
[10pt]
\displaystyle \rho_f\,\frac{\partial\,v_s}{\partial\,t} + \rho_w\,\frac{\partial\,w}{\partial\,t} + \frac{\eta}{\kappa}\,\frac{1}{\sqrt{\Omega}}\,(D+\Omega)^{1/2}\,w = -\frac{\partial\,p}{\partial\,x}+f_f,\label{LCBiot_w}\\
[10pt]
\displaystyle \sigma = (\lambda_f + 2\,\mu)\,\varepsilon - m\,\beta\,\xi,\label{LCBiot_sigma}\\
[10pt]
\displaystyle p = m\,(-\beta\,\varepsilon+\xi).\label{LCBiot_p}
\end{subnumcases}
We rearrange this system by separating  
$\frac{\partial\,v_s}{\partial\,t}$ and $\frac{\partial\,w}{\partial\,t}$ in (\ref{LCBiot_v}) and (\ref{LCBiot_w}) and using the definitions of $\varepsilon$ and $\xi$. Taking
\begin{equation}
\gamma = \frac{\eta}{\kappa}\,\frac{\rho}{\chi}\,\frac{1}{\sqrt{\Omega}},
\label{coef_gamma_omega_biot}
\end{equation}
one obtains the following system of equations of evolution 
\begin{subnumcases}{\label{S1}}
\displaystyle \frac{\partial\,v_s}{\partial\,t} - \frac{\rho_w}{\chi}\,\frac{\partial\,\sigma}{\partial\,x} - \frac{\rho _f}{\chi}\,\frac{\partial\,p}{\partial\,x} = \frac{\rho _f}{\rho}\,\gamma\,(D+\Omega)^{1/2}w+f_{v_s},\label{S1_1}\\
[15pt]
\displaystyle \frac{\partial\,w}{\partial\,t} + \frac{\rho _f}{\chi}\,\frac{\partial\,\sigma}{\partial\,x} + \frac{\rho }{\chi}\,\frac{\partial\,p}{\partial\,x} = -\,\gamma\,(D+\Omega)^{1/2}w+f_w,\label{S1_2}\\
[15pt]
\displaystyle \frac{\partial\,\sigma}{\partial\,t} - (\lambda_f + 2\mu)\,\frac{\partial\,v_s}{\partial\,x} - m\,\beta \,\frac{\partial\,w}{\partial\,x}=f_\sigma,\label{S1_3}\\
[15pt]
\displaystyle \frac{\partial\,p}{\partial\,t} + m\,\beta \, \frac{\partial\,v_s}{\partial\,x} + m\, \frac{\partial\,w}{\partial\,x}=f_p,\label{S1_4}
\end{subnumcases}
with $f_{v_s}=(\rho_w\,f_b-\rho_f\,f_f)\,/\,\chi$ and $f_w=(\rho\,f_f-\rho_f\,f_b)\,/\,\chi$. Terms $f_\sigma$ and $f_p$ have also been added to the derivatives of constitutive laws to simulate sources of mass.


\subsection{The diffusive representation}\label{SecPhysDR}

Taking
\begin{equation}
D_{\Omega}w(x,t) = \frac{\partial \,w}{\partial \,t}+\Omega\,w,
\end{equation}
the shifted fractional derivative (\ref{F_t}) can be written \cite{DUBOIS10}
\begin{equation}
(D+\Omega )^{1/2}w(x,t) = \frac{1}{\sqrt{\pi}}\,\int _0 ^t \frac{e^{-\Omega (t-\tau)}}{\sqrt{t-\tau}}D_{\Omega}w(x,\tau)\,d\tau.
\label{Dfrac}
\end{equation}
The operator $(D+\Omega)^{1/2}$  is not local in time and involves the entire time history of $w$. As we will see in section \ref{SecDA}, a different way of writing this derivative is more convenient for numerical evaluation. Based on Euler's $\Gamma$ function, the diffusive representation of the totally monotone function $\frac{1}{\sqrt{t}}$ \cite{DESCH88,MATIGNON10,THESE_HELE,STAFFANS94} is
\begin{equation}
\displaystyle \frac{1}{\sqrt{t}} = \frac{1}{\sqrt{\pi}} \int _0 ^\infty\,\frac{1}{\sqrt{\theta}}\,e^{-\theta t}d\theta .
\label{fonction_diffu}
\end{equation}
Substituting (\ref{fonction_diffu}) into (\ref{Dfrac}) gives
\begin{equation}
\begin{array}{ll}
\displaystyle (D+\Omega )^{1/2}w(x,t) & \displaystyle = \frac{1}{\pi}\,\int _0 ^t \int _0 ^{\infty}\frac{1}{\sqrt{\theta}}\,e^{-\theta (t-\tau)}\,e^{-\Omega (t-\tau)}\,D_{\Omega}w(x,\tau)\,d\theta\, d\tau,\\
[20pt]
& \displaystyle = \frac{1}{\pi}\, \int _0 ^{\infty}\frac{1}{\sqrt{\theta}}\,\psi (x,\theta,t)\,d\theta,
\end{array}
\label{derivee_frac}
\end{equation}
where the diffusive variable is defined as
\begin{equation}
\psi (x,\theta,t)=\int_0^t e^{-(\theta + \Omega )(t-\tau)}\,D_{\Omega}w(x,\tau )\,d\tau.
\label{variable_diffu}
\end{equation}
For the sake of clarity, the dependence on $\Omega$ and $w$ is omitted in $\psi$. From (\ref{variable_diffu}), it follows that the diffusive variable $\psi$ satisfies the ordinary differential equation
\begin{equation}
\left\lbrace 
\begin{array}{l}
\displaystyle \frac{\partial\,\psi}{\partial\,t} = -(\theta + \Omega )\,\psi + D_{\Omega}w, \\
[10pt]
\displaystyle \psi (x,\theta,0) = 0.
\end{array}
\right.
\label{EDO_psi}
\end{equation}
The diffusive representation therefore transforms a non-local problem (\ref{Dfrac}) into a continuum of local problems (\ref{EDO_psi}). It should be emphasized at this point that no approximations have been made up to now. The computational advantages of the diffusive representation will be seen in sections \ref{SecDA} and \ref{SecExp}, where the discretization of (\ref{derivee_frac}) and (\ref{EDO_psi}) will yield a tractable formulation.


\subsection{Energy of Biot-JKD}\label{SecPhysNRJ}

Now, we express the energy of the Biot-JKD model \ref{LCBiot}). This result generalizes the analysis performed in the LF range in \cite{THESE_EZZIANI}.

\begin{proposition}
Let
$$
E=E_1+E_2+E_3,
$$
with
\begin{equation}
\begin{array}{lll}
E_1  & = & \displaystyle \frac{\textstyle 1}{\textstyle 2}\,\int_{\mathbb{R}}\left(\rho\,v^2+\rho_w\,w^2+2\,\rho_f\,v\,w\right) dx\mbox{,} \\ 
[20pt]
E_2 & = & \displaystyle \frac{\textstyle 1}{\textstyle 2}\,\int_{\mathbb{R}}\left( \frac{\textstyle 1}{\textstyle C}\,\left(\sigma+\beta\,p\right)^2 + \frac{\textstyle 1}{\textstyle m}\,p^2\right)\,dx\mbox{,}\\
[20pt]
E_3 & = & \displaystyle \frac{\textstyle 1}{\textstyle 2}\,\int_{\mathbb{R}}\int_{\theta\in\mathbb{R}^+}\frac{\eta}{\kappa}\,\frac{1}{\pi}\,\frac{1}{\sqrt{\Omega\,\theta}}\,\frac{1}{\theta+2\,\Omega}\,(w-\psi)^2\,d\theta\,dx.
\end{array}
\label{EEnergie}
\end{equation}
Then $E$ is an energy which satisfies
\begin{equation}
\frac{\textstyle dE}{\textstyle dt} = -\int_{\mathbb{R}}\int_{\theta\in\mathbb{R}^+}\frac{\eta}{\kappa}\,\frac{1}{\pi}\,\frac{1}{\sqrt{\Omega\,\theta}}\,\frac{1}{\theta+2\,\Omega}\,\left( \Omega\,w^2+(\theta+\Omega)\,\psi^2\right) d\theta\,dx \;\leq\; 0.
\label{EdEdt}
\end{equation}
\label{PropositionNRJ}
\end{proposition}

Proposition \ref{PropositionNRJ} is proven in appendix 1. It calls for the following comments:
\begin{itemize}
\item the Biot-JKD model is well-posed;
\item when the viscosity of the saturating fluid is neglected ($\eta=0$), the energy of the system is conserved;
\item the terms in (\ref{EEnergie}) have a clearly physical significance: $E_1$ is the kinetic energy, and $E_2$ is the potential energy. The term $E_3$ corresponds to the kinetic energy resulting from the filtration velocity.
\end{itemize}


\subsection{Dispersion analysis}

\begin{figure}[htbp]
\begin{center}
\begin{tabular}{cc}
phase velocity of the fast wave & phase velocity of the slow wave\\
\includegraphics[scale=0.35]{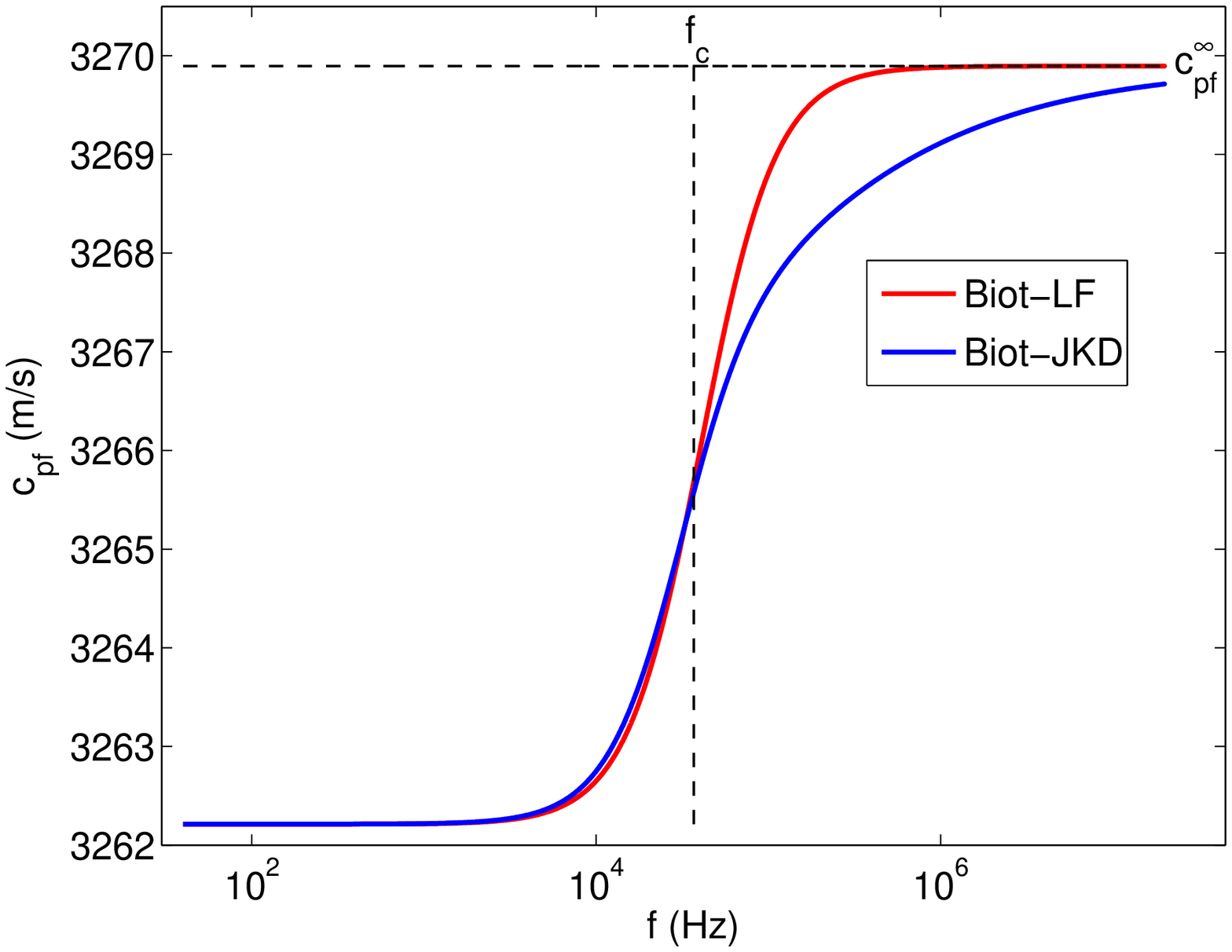} &
\includegraphics[scale=0.35]{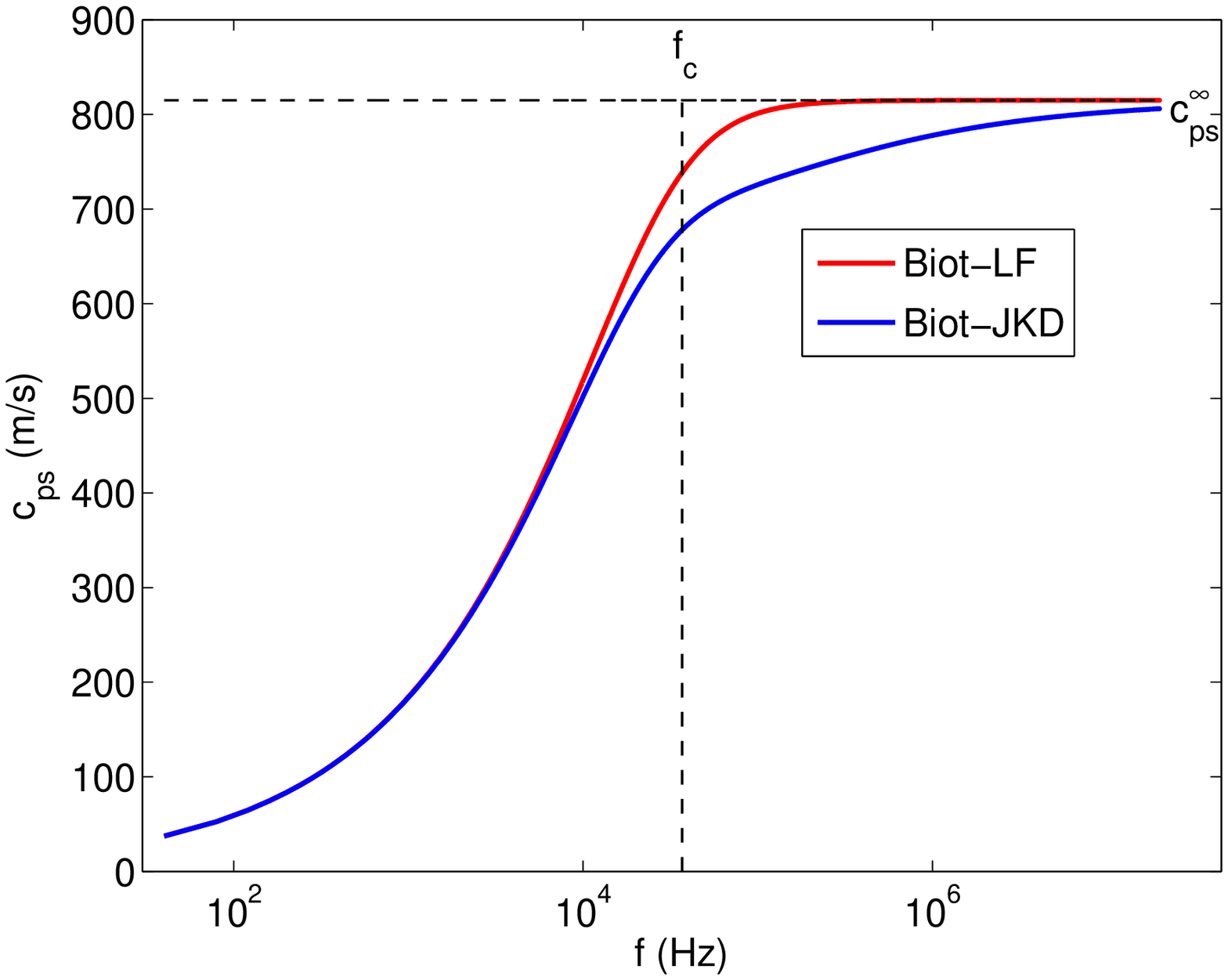}\\
attenuation of the fast wave & attenuation of the slow wave\\
\includegraphics[scale=0.35]{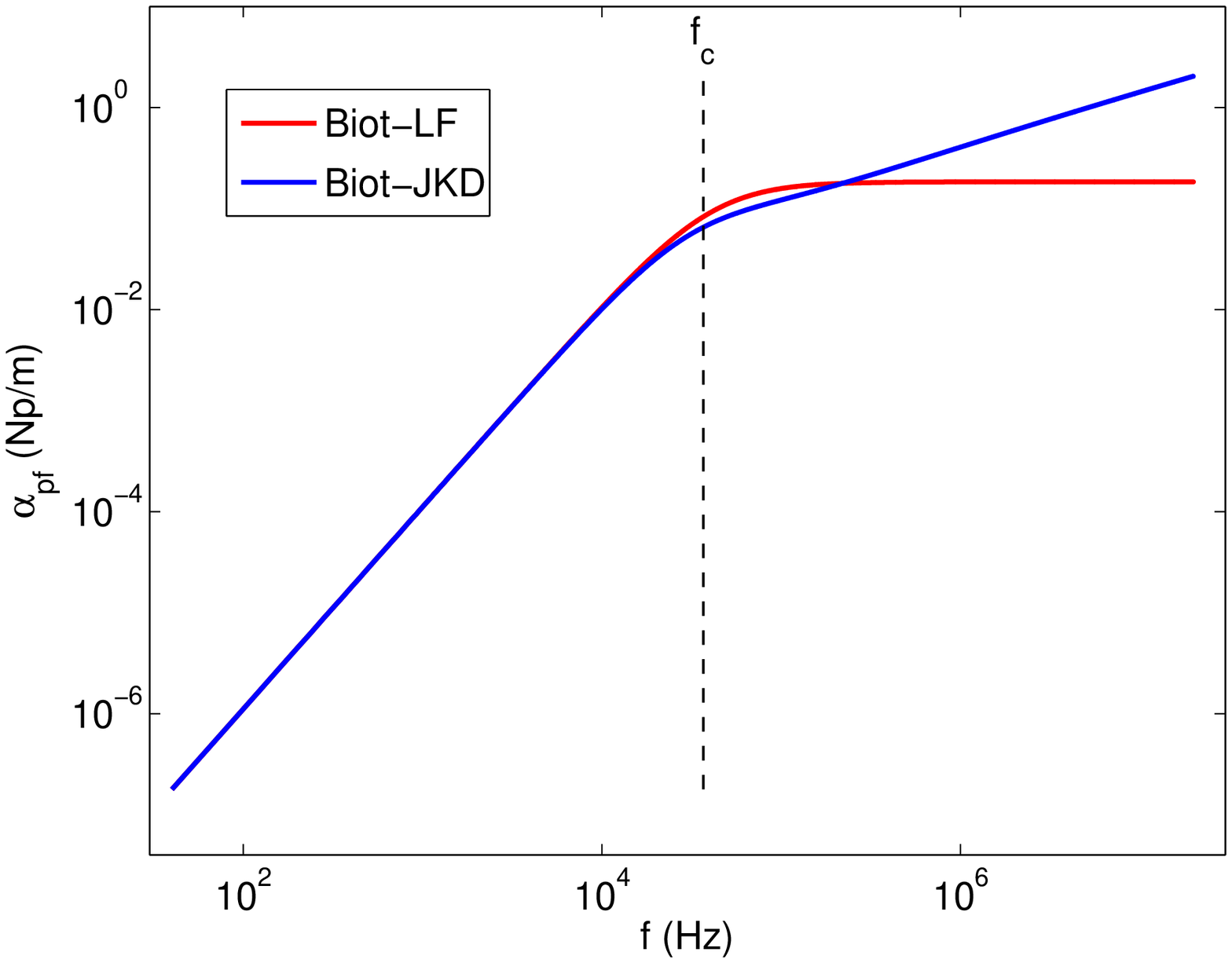} &
\includegraphics[scale=0.35]{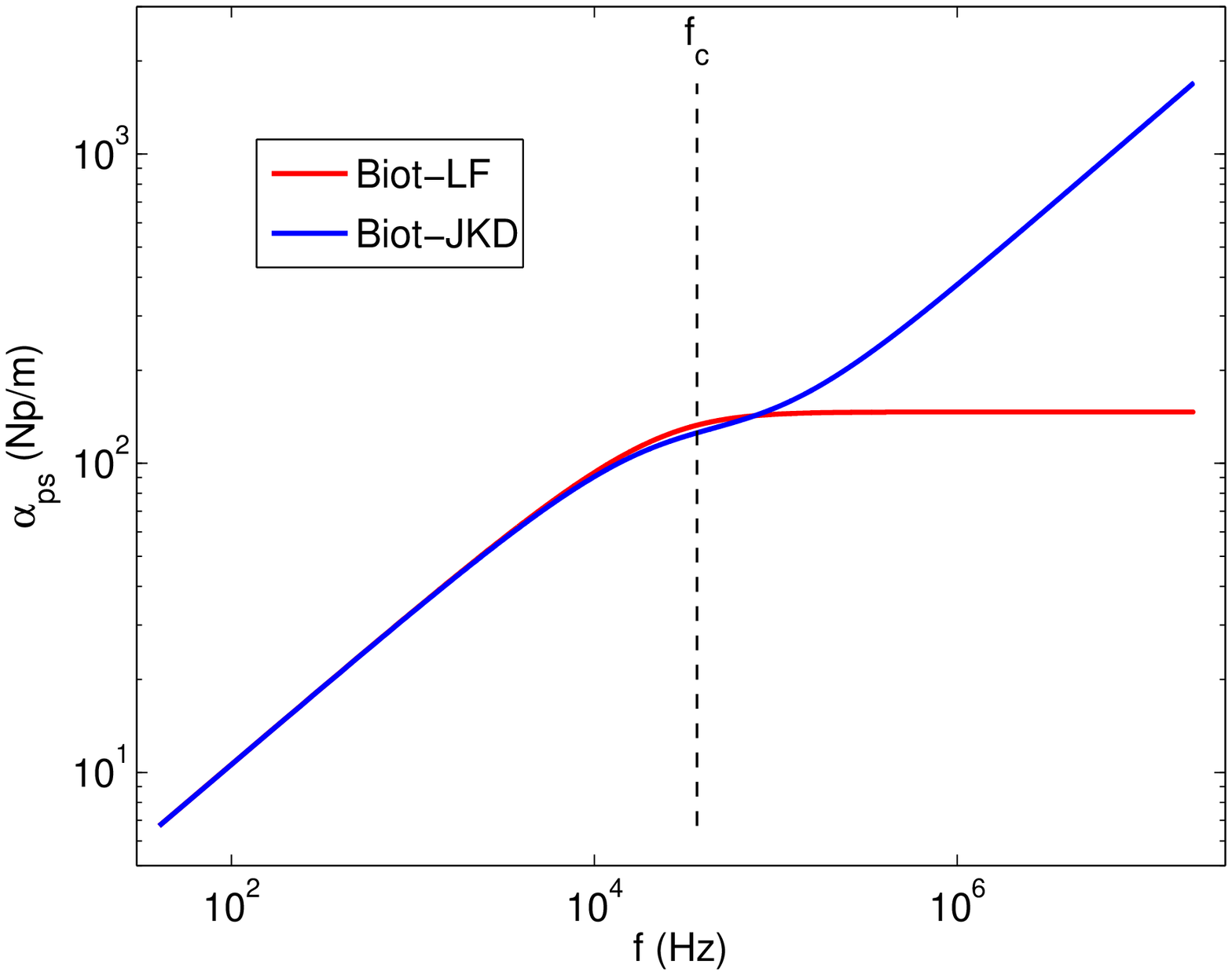}\\
\end{tabular}
\end{center}
\caption{Dispersion curves: comparison between Biot-LF and Biot-JKD.}
\label{dispersion_bf}
\end{figure}

Injecting a mode $e^{i(\omega t-kx)}$ in (\ref{S1}) gives the dispersion relation between the angular frequency $\omega$ and the wavenumber $k$. Taking
\begin{equation}
\left\lbrace \begin{array}{ll}
\displaystyle D_4 & = m\,(\lambda_0+2\,\mu)\mbox{,}\\
[12pt]
\displaystyle D_2(\omega) & = \displaystyle -\left( (\lambda_f+2\,\mu)\,\rho_w + m\,(\rho-2\,\rho_f\,\beta)\right) \,\omega^2 + i\,\omega\,\frac{\eta}{\kappa}\,\widehat{F}(\omega)\,(\lambda_f+2\,\mu)\mbox{,}\\
[10pt]
\displaystyle D_0(\omega) & = \displaystyle \chi\,\omega^4 - i\,\omega^3\,\frac{\eta}{\kappa}\,\rho\,\widehat{F}(\omega)\mbox{.}
\end{array}
\right. 
\label{coef_relation_dis}
\end{equation}
the dispersion relation takes the form
\begin{equation}
D_e(k,\omega ) = D_4\,k^4 + D_2(\omega)\,k^2 + D_0(\omega) = 0.
\label{relation_dispersion}
\end{equation}
Expressions (\ref{coef_relation_dis})-(\ref{relation_dispersion}) are valid in the case of both the Biot-LF and Biot-JKD  models with the frequency correction defined by
\begin{subnumcases}{\label{fonction_correction}
\widehat{F}(\omega ) = }
\displaystyle \widehat{F}_{LF}(\omega) = 1 & Biot-LF,\label{fonction_correction_BF}\\
[10pt]
\displaystyle \widehat{F}_{JKD}(\omega) = \displaystyle \frac{1}{\sqrt{\Omega}}\,(\Omega +i\,\omega)^{1/2} & Biot-JKD.\label{fonction_correction_JKD}
\end{subnumcases}
The solutions $k_{pf}$ and $k_{ps}$ of (\ref{relation_dispersion}) give the phase velocities $c_{pf} = \omega/\Re\mbox{e}(k_{pf})$ of the fast wave and $c_{ps} = \omega/\Re\mbox{e}(k_{ps})$ of the slow wave, with $0<c_{ps}<c_{pf}$. The attenuations $\alpha_{pf} = -\Im\mbox{m}(k_{pf})$ and $\alpha_{ps} = -\Im\mbox{m}(k_{ps})$ can also be deduced. Both the phase velocities and the attenuations of Biot-LF and Biot-JKD are strictly increasing functions of the frequency. 
The high frequency limits of fast and slow phase velocities, $c_{pf}^{\infty}$ and $c_{ps}^{\infty}$,  which are obtained by diagonalizing the left-hand side of system (\ref{S1}), satisfy the relation
\begin{equation}
\chi\,c^4 - \left( (\lambda_f+2\,\mu)\,\rho_w+m\,(\rho-2\,\rho_f\,\beta) \right)  \,c^2 + m\,(\lambda_0+2\,\mu) = 0.
\label{poly_cara_A}
\end{equation}
Figure \ref{dispersion_bf} shows the dispersion curves corresponding to the Biot-LF and Biot-JKD models. The physical parameters are those used in the numerical experiments presented in section \ref{SecExp}. Note that the scales are radically different in the case of fast and slow waves. The following properties can be observed:
\begin{itemize}
\item when $f<f_c$, the Biot-JKD and Biot-LF dispersion curves are very similar as might be expected, since $\displaystyle{\lim_{\omega \rightarrow 0} \widehat{F}_{JKD}(\omega)=1}$;
\item the fast wave is almost not affected by the frequency correction $\widehat{F}(\omega)$ while the slow wave is greatly affected;
\item when $f\ll f_c$, the slow wave degenerates to a diffusion process and is characterized by $\Re\mbox{e}(k_{ps})=\Im\mbox{m}(k_{ps})$. When $f>f_c$, the slow wave propagates but is greatly attenuated.
\end{itemize}

\section{The Biot-DA (diffusive approximation) model}\label{SecDA}

The aim of this section is to approximate the Biot-JKD model, using a numerically tractable approach. For this purpose, we follow a diffusive representation of fractional derivatives, initally proposed in \cite{HANYGA05}.

\subsection{The Biot-DA first-order system}\label{SecDAedp}

Using a quadrature formula on $N$ points, with weights $a_{\ell}$ and abscissa $\theta_{\ell}>0$, the diffusive representation (\ref{derivee_frac}) can be approximated by
\begin{equation}
\begin{array}{ll}
\displaystyle (D+\Omega )^{1/2}w(x,t) & \displaystyle = \frac{1}{\pi}\,\int _0 ^{\infty}\frac{1}{\sqrt{\theta}}\,\psi (x,t,\theta)\,d\theta, \\
[12pt]
& \displaystyle \simeq \sum \limits _{\ell=1} ^{N}a_{\ell} \,\psi (x,t,\theta_{\ell}),\\
[12pt]
& \displaystyle \equiv \sum \limits _{\ell=1} ^{N}a_{\ell} \,\psi _{\ell} (x,t).
\label{derivee_frac_rd}
\end{array}
\end{equation}
>From (\ref{EDO_psi}), the $N$ diffusive variables $\psi_{\ell}$ satisfy the ordinary differential equations
\begin{equation}
\left\lbrace 
\begin{array}{l}
\displaystyle \frac{\partial\,\psi_{\ell}}{\partial\,t} = -(\theta_{\ell} + \Omega )\,\psi_{\ell} + D_{\Omega}w, \\
[10pt]
\displaystyle \psi_{\ell} (x,0) = 0.
\end{array}
\right.
\label{EDO_psi_l}
\end{equation}
The fractional derivatives  are replaced by their diffusive approximation (\ref{derivee_frac_rd}) in the JKD model (\ref{S1}). Upon adding the equations (\ref{EDO_psi_l}) and performing some straightforward operations, the Biot-DA system is written as a first-order system in time and space 
\begin{equation}
\left\lbrace 
\begin{array}{l}
\displaystyle \frac{\partial\,v_s}{\partial\,t} - \frac{\rho _w}{\chi}\,\frac{\partial\,\sigma}{\partial\,x} - \frac{\rho _f}{\chi}\,\frac{\partial\,p}{\partial\,x} = \frac{\rho _f}{\rho}\,\gamma\,\sum \limits _{\ell =1}^{N}a_{\ell } \,\psi _{\ell }+f_{v_s},\\
[15pt]
\displaystyle \frac{\partial\,w}{\partial\,t} + \frac{\rho _f}{\chi}\,\frac{\partial\,\sigma}{\partial\,x} + \frac{\rho }{\chi}\,\frac{\partial\,p}{\partial\,x} = -\gamma\,\sum \limits _{\ell =1}^{N}a_{\ell }\, \psi _{\ell }+f_w,\\
[15pt]
\displaystyle \frac{\partial\,\sigma}{\partial\,t} - (\lambda_f + 2\mu)\,\frac{\partial\,v_s}{\partial\,x} - m\,\beta \,\frac{\partial\,w}{\partial\,x} = f_\sigma,\\
[15pt]
\displaystyle \frac{\partial\,p}{\partial\,t} + m\,\beta \,\frac{\partial\,v_s}{\partial\,x} + m \,\frac{\partial\,w}{\partial\,x} = f_p,\\
[15pt]
\displaystyle \frac{\partial\,\psi _j}{\partial\,t} + \frac{\rho _f}{\chi}\,\frac{\partial\,\sigma}{\partial\,x} + \frac{\rho }{\chi}\,\frac{\partial\,p}{\partial\,x} = \Omega \,w - \gamma\,\sum \limits _{\ell =1}^{N}a_{\ell}\, \psi _{\ell} -(\theta_j + \Omega ) \,\psi _j+f_w,\,\, j = 1,...,N\mbox{.}
\end{array}
\right. 
\label{syst2}
\end{equation}
Taking the vector of unknowns
\begin{equation}
{\bf U} = (v_s,w,\sigma,p,\psi _1,\ldots,\psi _N)^T
\end{equation}
and the source vector
\begin{equation}
{\bf F} = (f_{v_s},\,f_w,\,f_\sigma,\,f_p,\,f_w,\ldots,\,f_w)^T,
\end{equation}
the system (\ref{syst2}) can be written 
\begin{equation}
\frac{\partial\,{\bf U}}{\partial\,t} +{\bf A}\,\frac{\partial\,{\bf U}}{\partial\,x} = -{\bf S}\,{\bf U} +{\bf F},
\label{syst_hyperbolique}
\end{equation}
where ${\bf A}$ is the $(N+4)^2$ propagation matrix
\begin{equation}
{\bf A} = \left( 
\begin{array}{cccc|ccc}
0 & 0 & \displaystyle -\frac{\rho _w}{\chi} & \displaystyle -\frac{\rho _f}{\chi} & 0 & ... & 0 \\
\rule[-5mm]{0mm}{12mm} 0 & 0 & \displaystyle \frac{\rho _f}{\chi} & \displaystyle \frac{\rho}{\chi} & \vdots & ... & \vdots \\
\rule[-5mm]{0mm}{5mm} -(\lambda _f + 2\mu ) & -m\,\beta & 0 & 0 & \vdots & ... & \vdots \\
\rule[-5mm]{0mm}{5mm} m\,\beta & m  & 0 & 0 & 0 & ... & 0 \\ \hline

\displaystyle
\rule[-5mm]{0mm}{12mm} 0 & 0 & \displaystyle \frac{\rho _f}{\chi} & \displaystyle \frac{\rho}{\chi} & 0 & ... & 0 \\
\vdots & \vdots & \vdots & \vdots & \vdots & ... & \vdots \\
\rule[-5mm]{0mm}{12mm}0 & 0 & \displaystyle \frac{\rho _f}{\chi} & \displaystyle \frac{\rho}{\chi} & 0 & ... & 0 \\ 
\end{array}
\right) ,
\label{matriceA}
\end{equation}
and ${\bf S}$ is the $(N+4)^2$ dissipation matrix
\begin{equation}
{\bf S} = \left( \begin{array}{cccc|cccc}
0 & 0 & 0 & 0 & \displaystyle -\frac{\rho _f}{\rho}\,\gamma\,a_1 & \displaystyle -\frac{\rho _f}{\rho}\,\gamma\,a_2 & ... & \displaystyle -\frac{\rho _f}{\rho}\,\gamma\,a_{N} \\
\rule[-5mm]{0mm}{12mm}0 & 0 & 0 & 0 & \displaystyle \gamma\,a_1 & \displaystyle \gamma\,a_2 & ... & \displaystyle \gamma\,a_{N} \\
\rule[-5mm]{0mm}{5mm}0 & 0 & 0 & 0 & 0 & ... & ... & 0 \\
\rule[-5mm]{0mm}{3mm}0 & 0 & 0 & 0 & 0 & ... & ... & 0 \\ \hline

\rule[-5mm]{0mm}{12mm}0 & -\Omega & 0 & 0 & \displaystyle \gamma\,a_1 + (\theta_1 + \Omega ) & \displaystyle \gamma\,a_2 & ... & \displaystyle \gamma\,a_{N} \\
\rule[-5mm]{0mm}{5mm}0 & -\Omega & 0 & 0 & \displaystyle \gamma\,a_1 & \displaystyle \gamma\,a_2 + (\theta_2 + \Omega ) & \ldots & \displaystyle \gamma\,a_{N} \\
\rule[-5mm]{0mm}{5mm}\vdots & \vdots & \vdots & \vdots & \vdots & \vdots & \ddots & \vdots \\
\rule[-5mm]{0mm}{5mm}0 & -\Omega & 0 & 0 & \displaystyle \gamma\,a_1 & \displaystyle \gamma\,a_2 & ... & \displaystyle \gamma\,a_N + (\theta_{N} + \Omega ) \\
\end{array}
\right).
\label{matriceS}
\end{equation}
The size of the system increases linearly with the number $N$ of diffusive variables.


\subsection{Properties}\label{SecDAProp}

Four properties of system (\ref{syst_hyperbolique}) are specified:
\begin{itemize}
\item the eigenvalues of ${\bf A}$ (\ref{matriceA}) are real: $0$ with multiplicity $N$, $\pm c_{pf}^{\infty}$ and $\pm c_{ps}^{\infty}$, where the latter satisfies (\ref{poly_cara_A}). The system (\ref{syst_hyperbolique}) is therefore hyperbolic;
\item since the eigenvalues and eigenvectors do not depend on the diffusive coefficients, they are the same in both the  Biot-DA and Biot-LF or Biot-JKD models. This is not so in the case of the method presented in \cite{MASSON10}, where the propagation matrix is modified to account for the fractional derivative;
\item the dispersion analysis is obtained in the case of the Biot-DA model by replacing $\widehat{F}$ by
\begin{equation}
\displaystyle \widehat{F}_{DA}(\omega) = \displaystyle \frac{\Omega + i\,\omega}{\sqrt{\Omega}}\,\sum \limits _{\ell=1}^{N}\frac{a_{\ell}}{\theta_{\ell} + \Omega + i\,\omega } 
\label{fonction_correction_AD}
\end{equation}
in equations (\ref{coef_relation_dis})-(\ref{relation_dispersion});
\item in line with proposition \ref{PropositionNRJ}, an energy analysis of (\ref{syst2}) is performed.
\end{itemize}

\begin{proposition}
Let
$$
E=E_1+E_2+E_3,
$$
with
\begin{equation}
\begin{array}{lll}
E_1  & = & \displaystyle \frac{\textstyle 1}{\textstyle 2}\int_{\mathbb{R}}\left(\rho\,v^2+\rho_w\,w^2+2\rho_f\,v\,w\right) dx\mbox{,} \\ 
[20pt]
E_2 & = & \displaystyle \frac{\textstyle 1}{\textstyle 2}\int_{\mathbb{R}}\left( \frac{\textstyle 1}{\textstyle C}\,\left(\sigma+\beta\,p\right)^2 + \frac{\textstyle 1}{\textstyle m}\,p^2\right)\,dx\mbox{,}\\
[20pt]
E_3 & = & \displaystyle \frac{\textstyle 1}{\textstyle 2}\,\int_{\mathbb{R}}\sum\limits_{\ell=1}^N\,\frac{\eta}{\kappa}\,\frac{1}{\sqrt{\Omega}}\,\frac{a_{\ell}}{\theta_{\ell}+2\,\Omega}\,(w-\psi_{\ell})^2\,dx.\\
\end{array}
\label{Energie-AD}
\end{equation}
Then $E$ satisfies
\begin{equation}
\frac{\textstyle dE}{\textstyle dt} = -\int_{\mathbb{R}}\sum\limits_{\ell=1}^N\,\frac{\eta}{\kappa}\,\frac{1}{\sqrt{\Omega}}\,\frac{a_{\ell}}{\theta_{\ell}+2\,\Omega}\,\left( \Omega\,w^2+(\theta_{\ell}+\Omega)\,\psi_{\ell}^2\right) \,dx.
\label{dEdt-AD}
\end{equation}
\label{PropositionNRJ-AD}
\end{proposition}

Since the proof is very similar in this case, it will not be repeated. The terms $E_1$ and $E_2$ are the same in both the Biot-DA and Biot-JKD models, whereas $E_3$ and the time evolution of $E$ differ;  in Biot-DA, the sign depends on  the  coefficients introduced into the diffusive approximation. The abscissas $\theta_\ell$ of the quadrature formula are positive, but no sign criterion is given a priori for the weights $a_\ell$. $E$ therefore cannot be said to be a decreasing energy, except in the obvious case where all the $a_\ell$ are positive.


\subsection{Determination of the Biot-DA parameters}\label{SecDACoeff} 

The $a_{\ell}$ and $\theta_{\ell}$ in (\ref{derivee_frac_rd}) now have to be determined. In \cite{HANYGA05}, the authors used a general Laguerre quadrature formulas. We have tried using this approach, but it gave poor results. Very large numbers of diffusive variables were required to approximate the Biot-JKD model accurately, resulting in a huge computational cost. In addition, the Biot-DA model based on Laguerre functions does not converge by construction towards Biot-LF when the frequency tends towards 0, which is neither satisfactory nor physically realistic. Lastly, the involved coefficients do not depend on the physical factors (parameters, source) involved, which partly explains the above two weaknesses.

A different method of determining the $2\,N$ coefficients $a_{\ell}$ and $\theta_{\ell}$ in the diffusive approximation (\ref{syst2}) is therefore used, in order to approach $\widehat{F}_{JKD}(\omega)$ (\ref{fonction_correction}) by $\widehat{F}_{DA}(\omega)$ (\ref{fonction_correction_AD}) in a given frequency range of interest. Let $Q(\omega)$ be the optimized quantity and $Q_{ref}(\omega)$ be the desired quantity
\begin{subnumcases}
\displaystyle Q(\omega) = \frac{\widehat{F}_{DA}(\omega)}{\widehat{F}_{JKD}(\omega)} = \sum \limits _{\ell=1}^{N}a_{\ell}\,\frac{(\Omega + i\,\omega)^{1/2}}{\theta_{\ell} + \Omega + i\,\omega } = \sum \limits _{\ell=1}^{N}a_{\ell}\,q_{\ell}(\omega),\label{Q_ql}\\
[15pt]
\displaystyle Q_{ref}(\omega) = 1.
\label{q_ref_opti}
\end{subnumcases}
We implement a linear optimization procedure \cite{EMMERICH87,GROBY06,LOMBARD11} in order to minimize the distance between $Q(\omega)$ and $Q_{ref}(\omega)$ in the interval $[\omega_{min},\omega_{max}]$ centered on $\omega_0 = 2\,\pi\,f_0$, where $f_0$ is the central frequency of the source. The abscissas $\theta_{\ell}$ are fixed and distributed linearly on a logarithmic scale
\begin{equation}
\theta_{\ell} = \omega_{min}\left( \frac{\omega_{max}}{\omega_{min}}\right) ^{\frac{\ell-1}{N-1}}\mbox{,}\qquad \ell = 1,...,N.
\end{equation}
The weights $a_{\ell}$ are obtained by solving the system
\begin{equation}
\sum \limits _{\ell=1}^{N}a_{\ell}\,q_{\ell}(\tilde{\omega}_k) = 1\mbox{,}\qquad k = 1,...,K,
\end{equation}
where the $\tilde{\omega}_k$ are also distributed linearly on a logarithmic scale of $K$ points
\begin{equation}
\tilde{\omega}_k = \omega_{min}\left( \frac{\omega_{max}}{\omega_{min}}\right) ^{\frac{k-1}{K-1}}\mbox{,}\qquad k = 1,...,K.
\label{omega_k}
\end{equation}
Since the $q_{\ell}(\omega)$ are complex functions, optimization is performed simultaneously on the real and imaginary parts 
\begin{equation}
\left\lbrace 
\begin{array}{ll}
\displaystyle \sum \limits _{\ell=1}^{N}a_{\ell}\,\mathbb{R}\mbox{e}(q_{\ell}(\tilde{\omega}_k))  & = 1,\\
[12pt]
\displaystyle \sum \limits _{\ell=1}^{N}a_{\ell}\,\mathbb{I}\mbox{m}(q_{\ell}(\tilde{\omega}_k))  & = 0\mbox{,}\qquad k=1,...,K.\\
\end{array}
\right. 
\label{syst_opti}
\end{equation}
A square system is obtained when $2K=N$, whereas $2\,K>N$ yields an overdetermined system, which can be solved by writing normal equations \cite{NRPAS}. For practical purposes, we use $\omega_{min}=\omega_0/10$ and $\omega_{max}=10\,\omega_0$, as in \cite{LOMBARD11}. 

\begin{figure}[htbp]
\begin{center}
\begin{tabular}{cc}
$2\,K=N$ & $K=N$ \\
\includegraphics[scale=0.35]{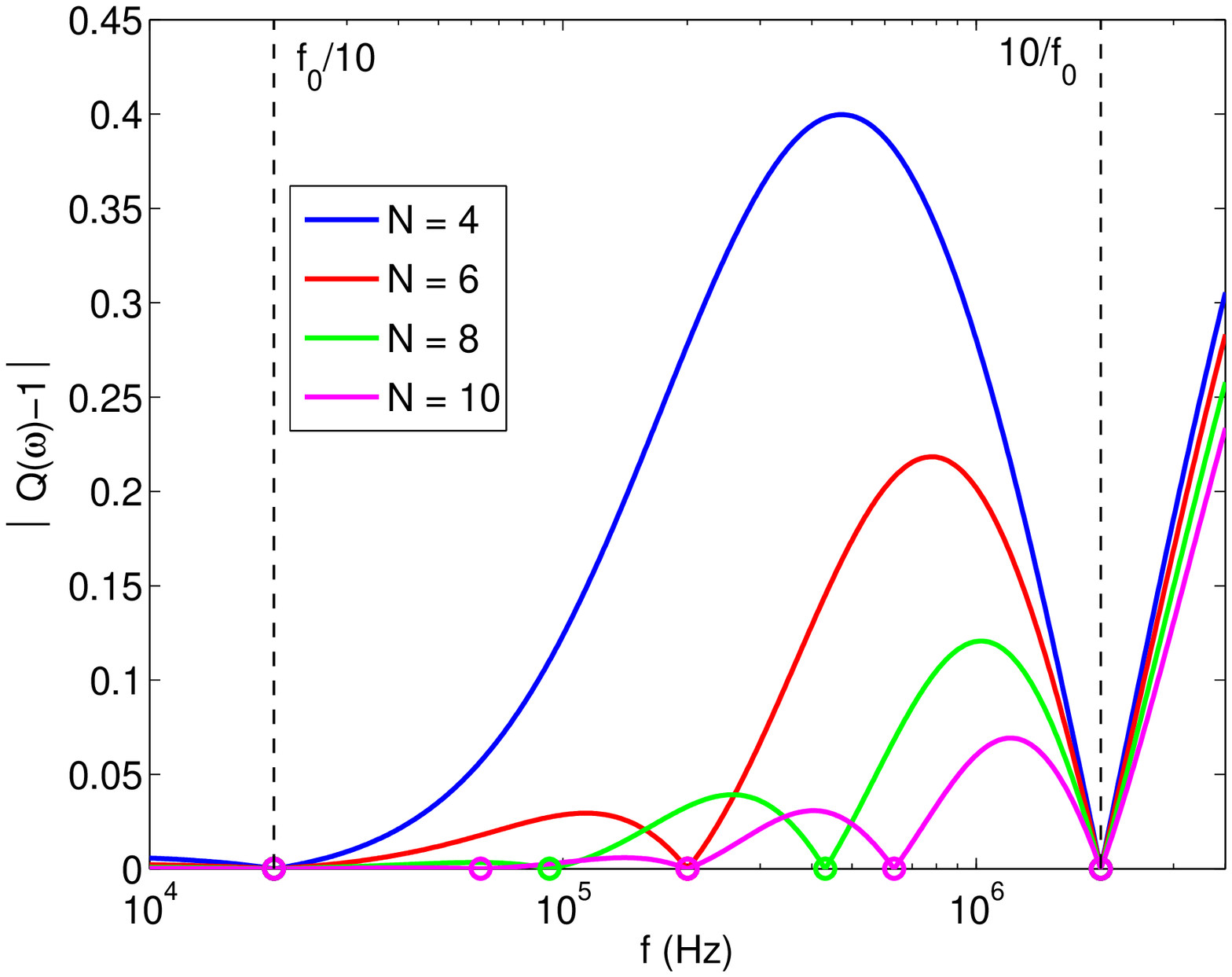} & 
\includegraphics[scale=0.35]{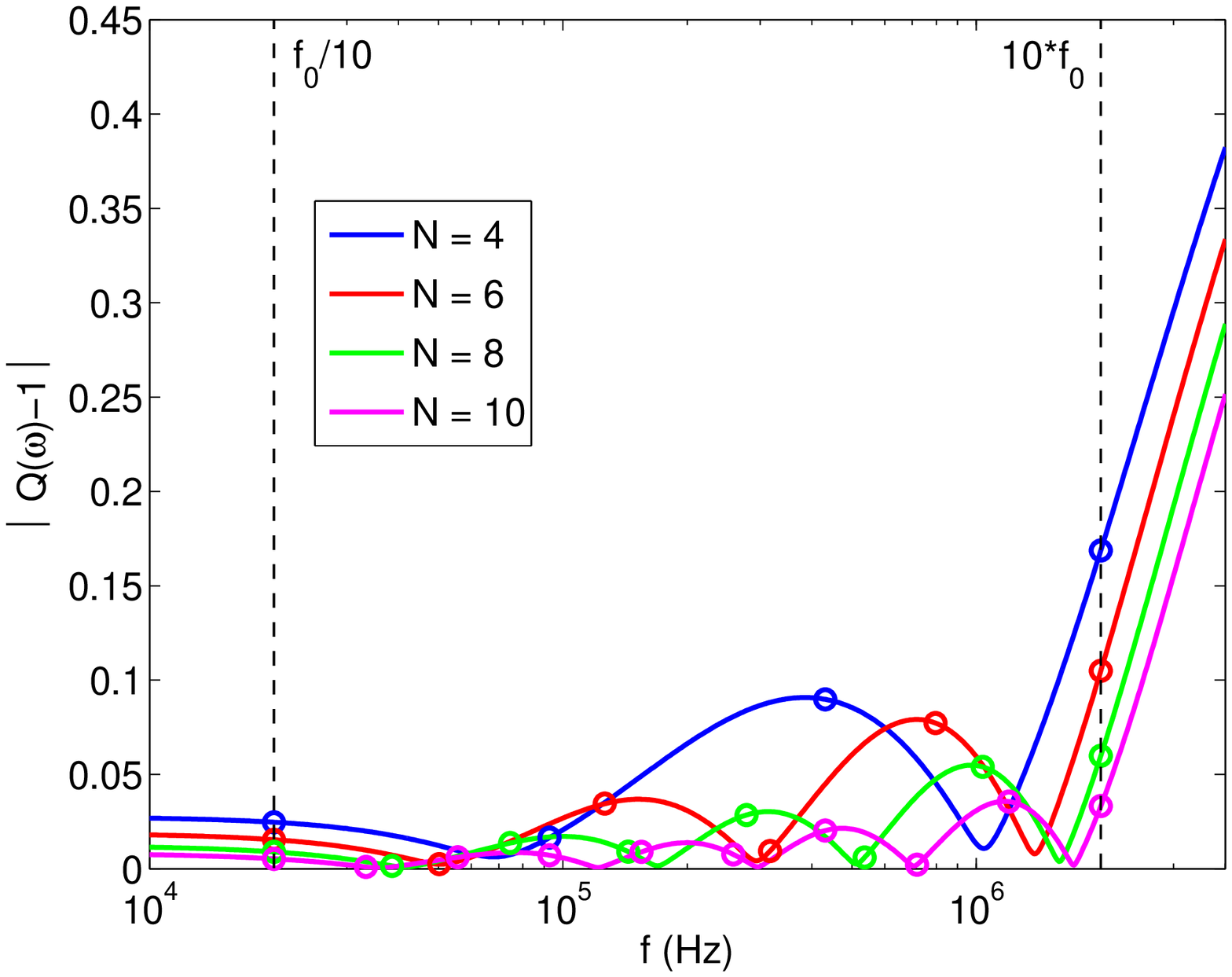}\\
$K=2N$ & $K=3N$ \\
\includegraphics[scale=0.35]{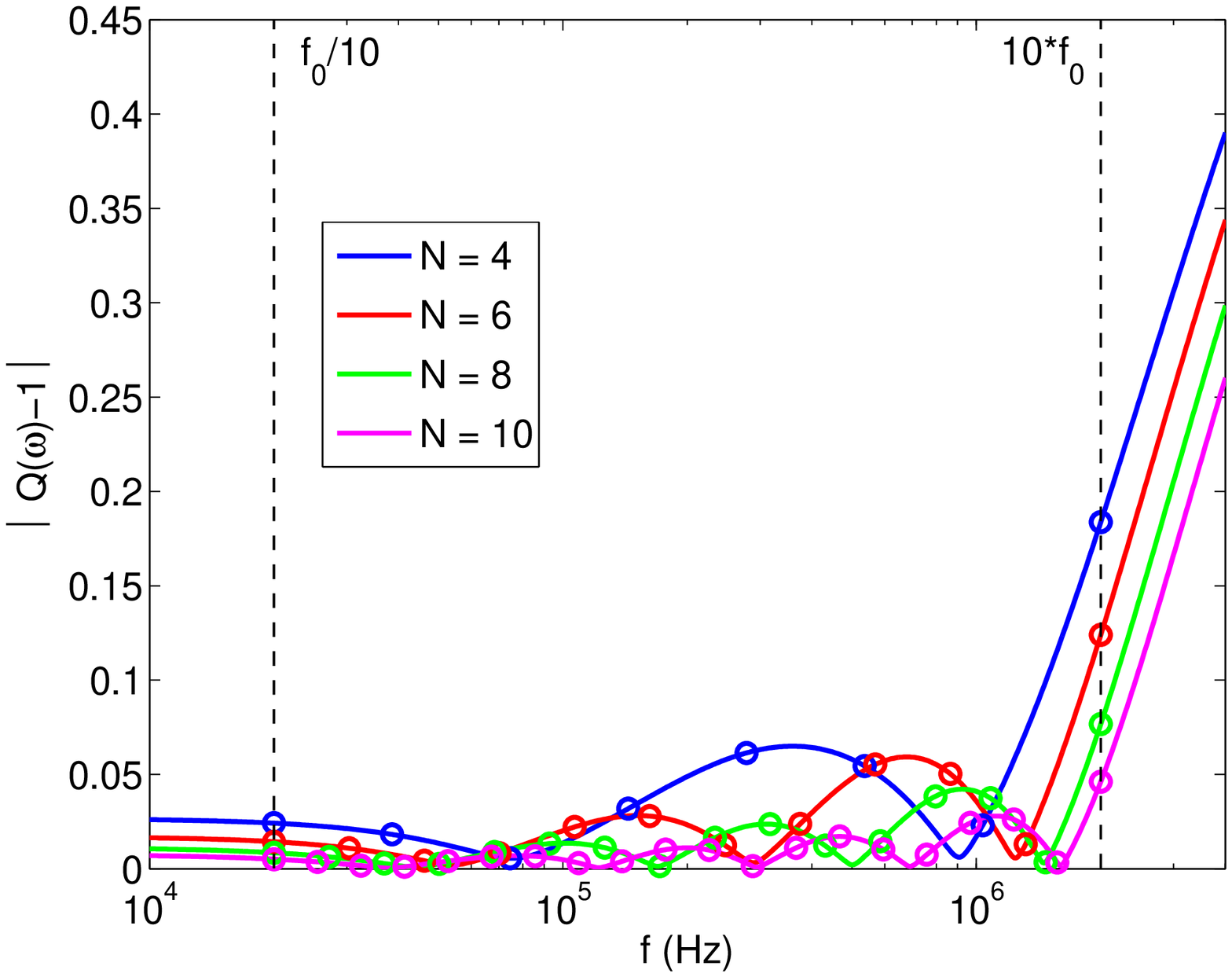} & 
\includegraphics[scale=0.35]{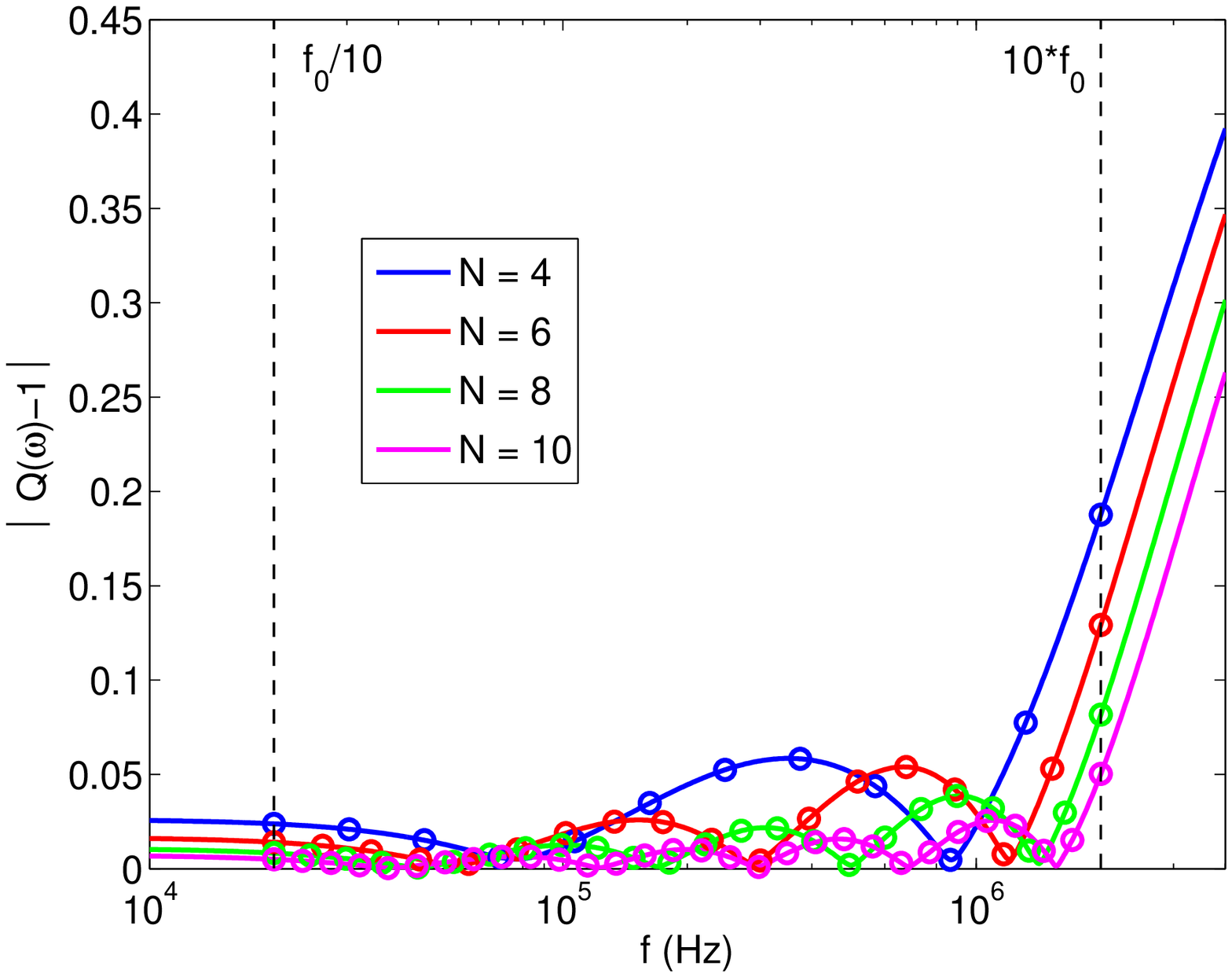}\\
\end{tabular}
\end{center}
\caption{Relative error $\mid Q(\omega)-1 \mid$ in (\ref{q_ref_opti}) in the case of various values of $(K,\,N)$.}
\label{fig_opti_erreur}
\end{figure}

\begin{figure}[htbp]
\begin{center}
\begin{tabular}{cc}
\includegraphics[scale=0.35]{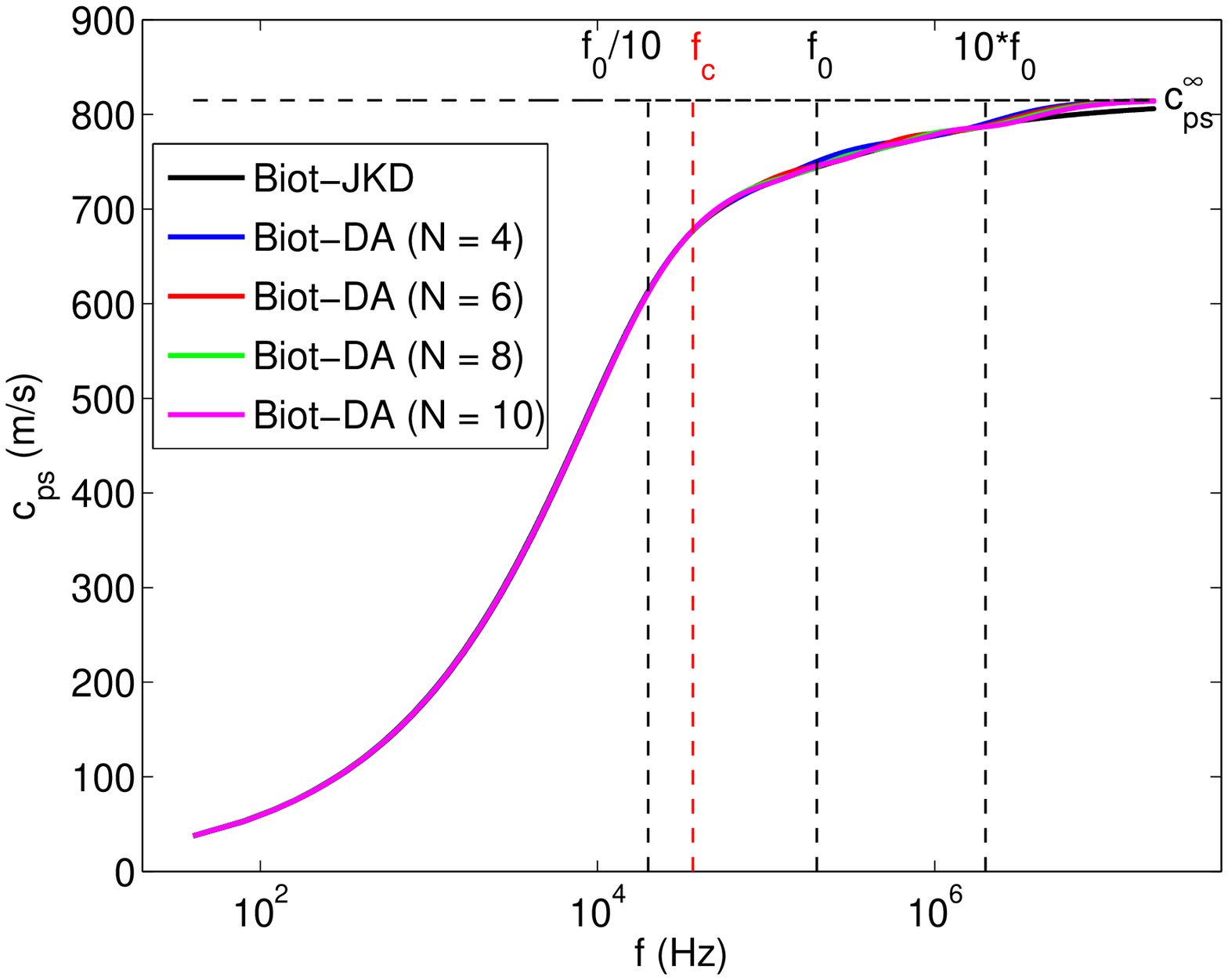} & 
\includegraphics[scale=0.35]{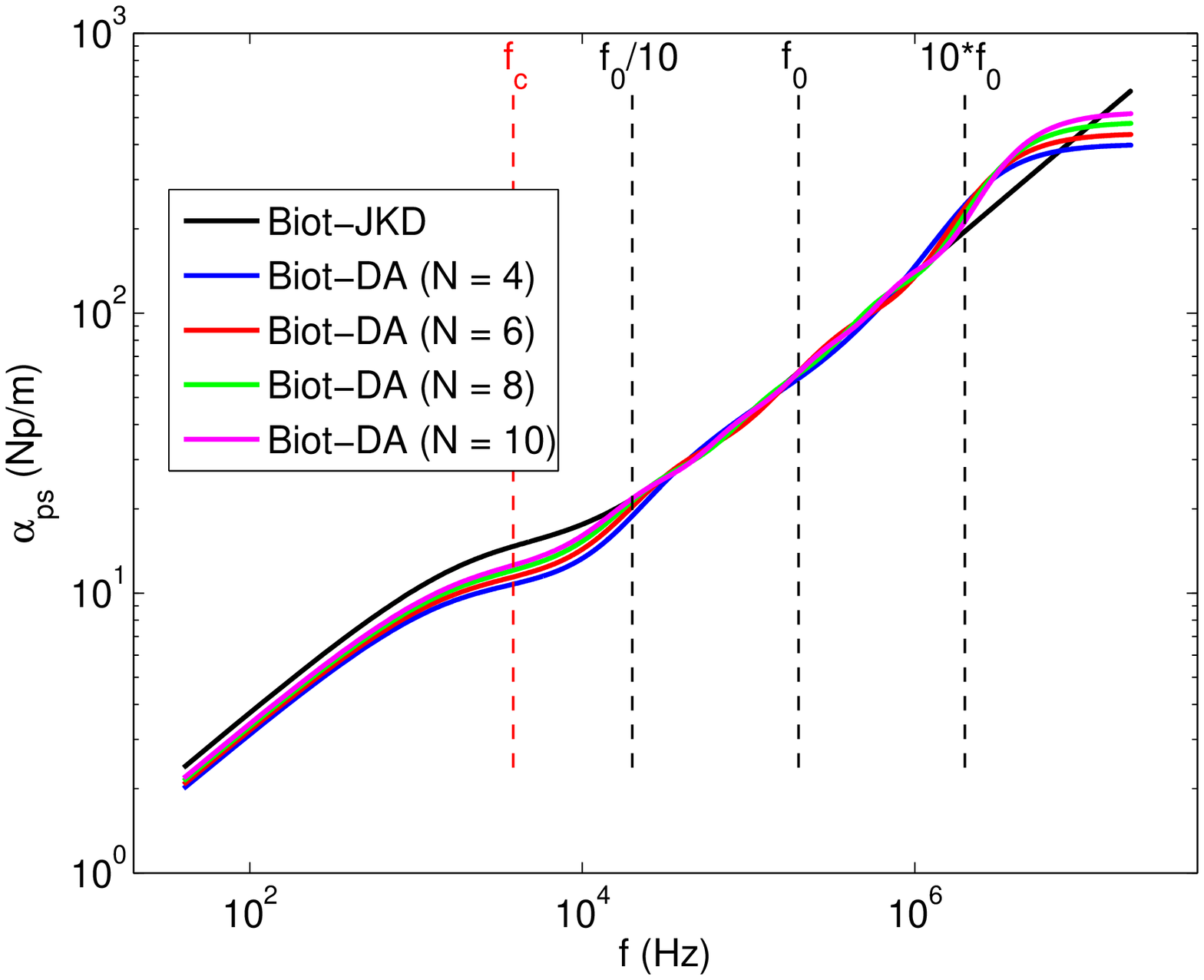}
\end{tabular}
\end{center}
\caption{Phase velocity $c_{ps}$ (left) and attenuation $\alpha_{ps}$ (right) of the slow wave obtained with the Biot-DA model, in terms of the number of diffusive variables.}
\label{fig_opti_disp}
\end{figure}

Figure \ref{fig_opti_erreur} illustrates the influence of $N$ and $K$ on the accuracy of the optimization procedure. As can be observed in this figure, the errors are smaller with the overdeterminated system ($K=N,2N,3N$) than with the square one. However, increasing the size of the system does not really improve the accuracy. In what follows, we will therefore always use the values $K=N$. The influence of the number of diffusive variables on the physical properties of the system is presented in figure \ref{fig_opti_disp}. We focus here on the slow wave, since it is more sensitive to the frequency correction. As was to be expected, the accuracy of the approximation of the Biot-JKD phase velocity and attenuation given by the Biot-DA model increases with $N$.

\begin{figure}[htbp]
\begin{center}
\begin{tabular}{cc}
\includegraphics[scale=0.35]{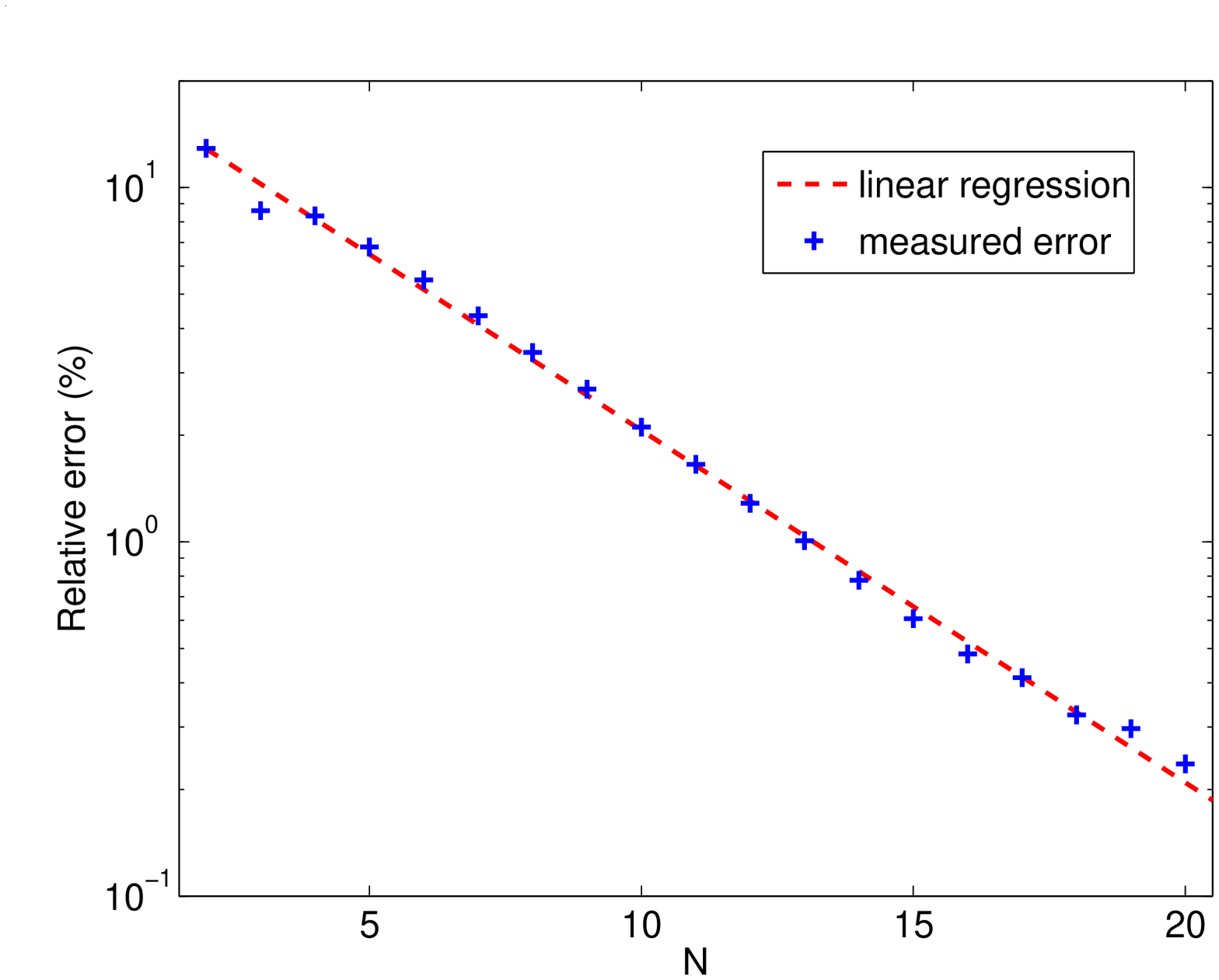}&
\includegraphics[scale=0.35]{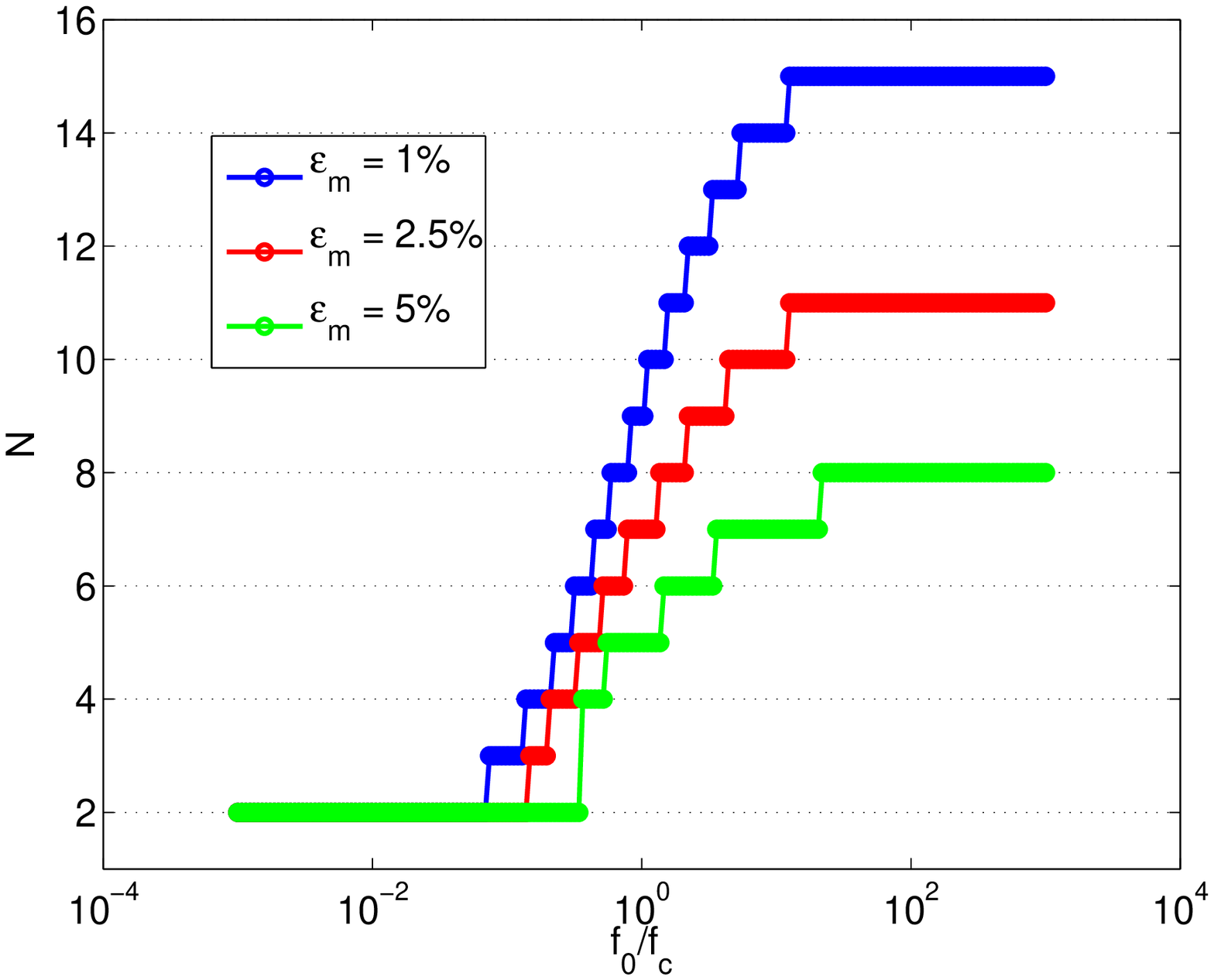}
\end{tabular}
\end{center}
\caption{Determining the number of diffusive variables $N$. Left: relative error $\varepsilon_m$ in terms of the number of $N$; the dashed line is proportional to $N^{-1}$. Right: required value of $N$ in terms of $f_0/f_c$ and the required accuracy $\varepsilon_m$ (b).}
\label{fig_opti_erreur_rel_N}
\end{figure}

To determine $N$ in terms of the required accuracy, $\varepsilon_m=||Q(\omega)-1||_{L_2}$ is measured in the frequency range of interest $[f_0/10, 10 f_0]$. This norm amounts to the relative error between $\widehat{F}_{DA}(\omega)$ and $\widehat{F}_{JKD}(\omega)$. With $N\leq 20$, this error is proportional to $N^{-1}$, as can be seen from figure \ref{fig_opti_erreur_rel_N}-(a). At larger values of $N$, the system is poorly conditioned and the order of convergence  deteriorates (not shown here); in practice, this is not penalizing, however, since large values of $N$ are of no use. An example of the parametric determination of $N$ in terms of the frequency range and the desired accuracy is also given in figure \ref{fig_opti_erreur_rel_N}-(b). In the following numerical tests, $N=6$ variables are used, giving the modeling error $\varepsilon_m \simeq 5.5\%$.

Lastly, the sign of weights $a_\ell$ was examined in a large number of configurations. In each case, some negative values were obtained with the linear optimization process (\ref{syst_opti}). As stated in proposition \ref{PropositionNRJ-AD}, the well-possedness of Biot-DA could not therefore be proved. A nonlinear optimization procedure with a positivity constraint was then applied \cite{NOCEDAL99}, but almost all the $a_\ell$ obtained were equal to zero.  In the numerical experiments, the negativity of some $a_\ell$ has never raised any problems. This question is addressed in detail at the end of section \ref{SecExpTest1}. 


\section{Numerical modeling}\label{SecNum}

\subsection{Splitting}\label{SecNumSpliiting}

In order to integrate the Biot-DA system (\ref{syst_hyperbolique}), a uniform grid is introduced, with mesh size $\Delta\,x$ and time step $\Delta\,t$. The approximation of the exact solution ${\bf U}(x_j = j\,\Delta\,x,t_n = n\,\Delta\,t)$ is denoted by ${\bf U}_j^n$. If an unsplit integration of (\ref{syst_hyperbolique}) is performed, Von-Neumann analysis typically yields the stability condition
\begin{equation}
\Delta\,t \leq \min \left( \Upsilon\,\frac{\Delta\,x}{c_{pf}^{\infty}} \; , \; \frac{2}{R({\bf S})}\right) ,
\label{CFL_direct}
\end{equation}
where $R({\bf S})$ is the spectral radius of ${\bf S}$, and $\Upsilon>0$ depends on the numerical scheme. We have no theoretical estimate of $R({\bf S})$, but numerical studies have shown that this value is similar to that of the spectral radius in LF: $\frac{\eta}{\kappa}\,\frac{\rho}{\chi}$, which can be very large \cite{CHIAVASSA11}. The time step can therefore be highly penalized in this case (\ref{CFL_direct}).

A more efficient strategy is adopted here, which consists in splitting the original system (\ref{syst_hyperbolique}) into a propagative part and a diffusive part (\ref{splitting_diffu})
\begin{numcases}{\label{SPLITTING}}
\displaystyle
\frac{\partial\,{\bf U}}{\partial\,t}+{\bf A}\,\frac{\partial\,{\bf U}}{\partial\,x} = 0,\label{split_propa}\\
[12pt]
\displaystyle
\frac{\partial\,{\bf U}}{\partial\,t} = -{\bf S}\,{\bf U}.\label{splitting_diffu}
\end{numcases}
For the sake of simplicity, the source term ${\bf F}$ has been omitted here. The discrete operators associated with steps (\ref{split_propa}) and (\ref{splitting_diffu}) are denoted by ${\bf H}_a$ and ${\bf H}_b$, respectively. The second-order Strang splitting \cite{LEVEQUE02} is then used to integrate (\ref{syst_hyperbolique}) between $t_n$ ant $t_{n+1}$, giving the  time-marching
\begin{equation}
\begin{array}{lllll}
\displaystyle
&\bullet& {\bf U}_{j}^{(1)}&=&{\bf H}_{b}(\frac{\Delta\,t}{2})\,{\bf U}_{j}^{n},\\
[6pt]
\displaystyle
&\bullet& {\bf U}_{j}^{(2)}&=&{\bf H}_{a}(\Delta\,t)\,{\bf U}_{j}^{(1)},\\
[6pt]
\displaystyle
&\bullet& {\bf U}_{j}^{n+1}&=&{\bf H}_{b}(\frac{\Delta\,t}{2})\,{\bf U}_{j}^{(2)}.
\end{array}
\label{AlgoSplitting}
\end{equation}
The discrete operator ${\bf H}_a$ associated with the propagative part (\ref{split_propa}) is an ADER 4 (Arbitrary DERivatives) scheme \cite{SCHWARTZKOPFF04}. This scheme is fourth-order accurate in space and time, is dispersive of order 4 and dissipative of order 6, and has a stability limit $\Upsilon=1$. On Cartesian grids, ADER 4 amounts to a fourth-order Lax-Wendroff scheme, and  can be written
\begin{equation}
\left\lbrace 
\begin{array}{l}
\displaystyle {\bf H}_a(\Delta\,t)\,{\bf U}_j^{(1)}= {\bf U}_j^{(1)} - \sum \limits _{s=-2}^{+2} {\bf C}_s \, {\bf U}_{j+s}^{(1)},\\
[12pt]
\displaystyle {\bf C}_s = -\sum _{m=1}^{4}\gamma_{m,s}\,\left( -{\bf A}\,\frac{\Delta\,t}{\Delta\,x}\right) ^m,
\end{array}
\right.
\label{schema_ADER}
\end{equation}
where the coefficients $\gamma_{m,s}$ are given in table \ref{coef_ader4}.

\begin{table}[htbp]
\begin{center}
\begin{small}
\begin{tabular}{l|cccc}
\rule[-3mm]{0mm}{8mm} $\gamma_{m,s}$ & $m=1$ & $m=2$ & $m=3$ & $m=4$\\ \hline
\rule[-3mm]{0mm}{8mm} $s=-2$  & $1/12$ & $1/24$ & $-1/12$ & $-1/24$ \\
\rule[-3mm]{0mm}{8mm} $s=-1$  & $-2/3$ & $-2/3$ & $1/6$ & $1/6$ \\
\rule[-3mm]{0mm}{8mm} $s=0$   & $0$ & $5/4$ & $0$ & $-1/4$ \\
\rule[-3mm]{0mm}{8mm} $s=+1$  & $2/3$ & $-2/3$ & $-1/6$ & $1/6$ \\
\rule[-3mm]{0mm}{8mm} $s=+2$  & $-1/12$ & $1/24$ & $1/12$ & $-1/24$ \\
\end{tabular}
\end{small}
\end{center}
\caption{Coefficients of the ADER 4 scheme.}
\label{coef_ader4}
\end{table} 

Since the physical parameters do not vary with time, the diffusive part  (\ref{splitting_diffu}) can be solved exactly. This gives
\begin{equation}
{\bf H}_b\left(\frac{\Delta\,t}{2}\right)\,{\bf U}_j = e^{-\frac{\Delta\,t}{2}\,{\bf S}}\,{\bf U}_j.
\label{split_diffu_exp}
\end{equation}
The matrix $e^{-\frac{\Delta\,t}{2}\,{\bf S}}$ is computed numerically using the $(r/q)$ Pad\'e approximation in the "scaling and squaring method" \cite{MOLER03}, which is given by the expression
\begin{equation}
\left\lbrace
\begin{array}{l}
\displaystyle e^{-\frac{\Delta\,t}{2}\,{\bf S}} \approx R_{rq}\left(-\frac{\Delta\,t}{2}\,{\bf S}\right) = \frac{N_{rq}(-\frac{\Delta\,t}{2}\,{\bf S})}{D_{rq}\left(-\frac{\Delta\,t}{2}\,{\bf S}\right)}, \\
[15pt]
\displaystyle N_{rq}\left(-\frac{\Delta\,t}{2}\,{\bf S}\right) = \sum \limits _{k=0}^r \frac{(r+q-k)\,!\,r\,!}{(r+q)\,!\,k\,!\,(r-k)\,!}\,\left(-\frac{\Delta\,t}{2}\,{\bf S}\right)^k,\\
[15pt]
\displaystyle D_{rq}\left(-\frac{\Delta\,t}{2}\,{\bf S}\right) = \sum \limits _{k=0}^q \frac{(r+q-k)\,!\,q\,!}{(r+q)\,!\,k\,!\,(r-k)\,!}\,\left(\frac{\Delta\,t}{2}\,{\bf S}\right)^k.\\
\end{array}
\right.
\end{equation}
In the following numerical experiments, the parameters $r=q=6$ are used.

It remains to verify that the numerical integration of the diffusive step (\ref{split_diffu_exp}) is unconditionally stable. This is achieved as follows.

\begin{proposition}
The diffusive part of the splitting (\ref{splitting_diffu}) is well-posed whatever the weights $a_{\ell}$ in the diffusive approximation (\ref{derivee_frac_rd}).
\label{PropositionStabExp}
\end{proposition}

Proposition \ref{PropositionStabExp} is proven in appendix 2. It follows that the solution of system (\ref{nrj_split}) is bounded and that the eigenvalues of $-\frac{\Delta\,t}{2}\,{\bf S}$ are then in the left half space. As a consequence, the $R_{qq}$ Pad\'e approximation is always stable \cite{MOLER03}. The full algorithm (\ref{AlgoSplitting}) is therefore stable under the optimum stability condition
\begin{equation}
\Delta\,t \leq  \Upsilon\,\frac{\Delta\,x}{c_{pf}^{\infty}},
\label{CFL_dsplit}
\end{equation} 
which is always independent of the Biot-DA model coefficients.


\section{Numerical experiments}\label{SecExp}

\subsection{General configuration}\label{SecExpConfig}

\begin{table}[htbp]
\begin{center}
\begin{tabular}{lll}
\hline
\rule[-1mm]{0mm}{5mm} Saturating fluid & $\rho_f$ (kg/m$^3$) & $1000$\\
\rule[-1mm]{0mm}{5mm} & $\eta$ (Pa.s)  & $10^{-3}$\\
\rule[-1mm]{0mm}{5mm} Grain & $\rho_s$ (kg/m$^3$) & $2644$\\
\rule[-1mm]{0mm}{5mm} & $\mu$ (Pa) & $7.04\,10^9$\\
\rule[-1mm]{0mm}{5mm} Matrix & $\phi$ & $0.2$\\
\rule[-1mm]{0mm}{5mm} & $a$ & $2.4$\\
\rule[-1mm]{0mm}{5mm} & $\kappa$ (m$^2$) & $3.6\,10^{-13}$\\
\rule[-1mm]{0mm}{5mm} & $\lambda_f$ (Pa) & $1.06\,10^{10}$\\
\rule[-1mm]{0mm}{5mm} & $m$ (Pa) & $9.70\,10^{9}$\\
\rule[-1mm]{0mm}{5mm} & $\beta$ & $0.720$\\
\rule[-1mm]{0mm}{5mm} & $\Lambda$ (m) & $5.88\,10^{-6}$\\ \hline
\rule[-1mm]{0mm}{5mm} Phase velocities & $c_{pf}^{\infty}$ (m/s) & $3269.89$\\
\rule[-1mm]{0mm}{6mm} & $c_{ps}^{\infty}$ (m/s) & $814.95$\\
\rule[-1mm]{0mm}{6mm} & $c_{pf}^{\infty}/c_{ps}^{\infty}$ & $4.01$\\
\rule[-1mm]{0mm}{6mm} & $f_c$ (Hz) & $3.68\,10^{4}$\\ \hline
\end{tabular}
\end{center}
\caption{Physical parameters used in numerical experiments.}
\label{para_phy}
\end{table}

\begin{figure}[htbp]
\begin{center}
\begin{tabular}{cc}
\includegraphics[scale=0.35]{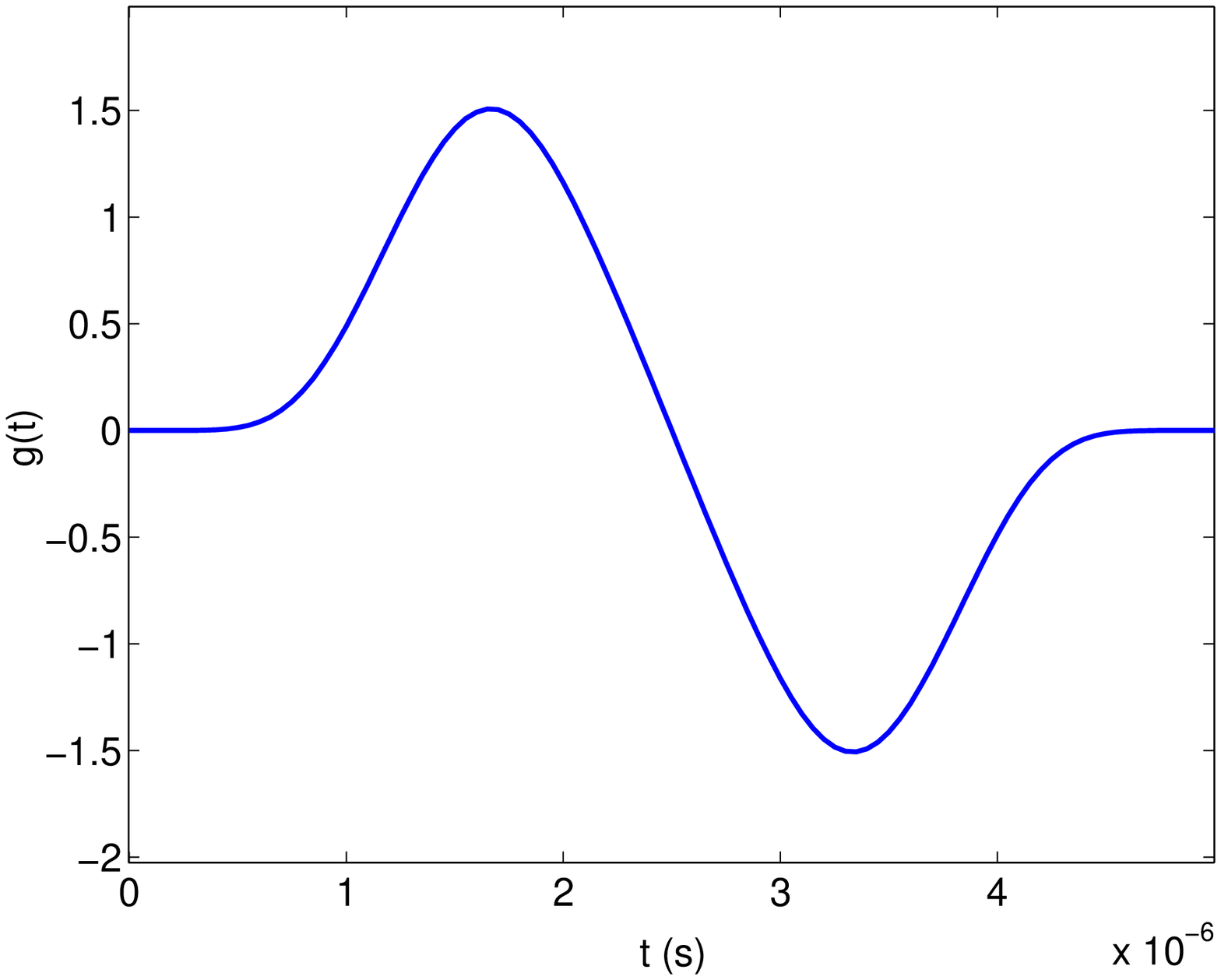} & 
\includegraphics[scale=0.35]{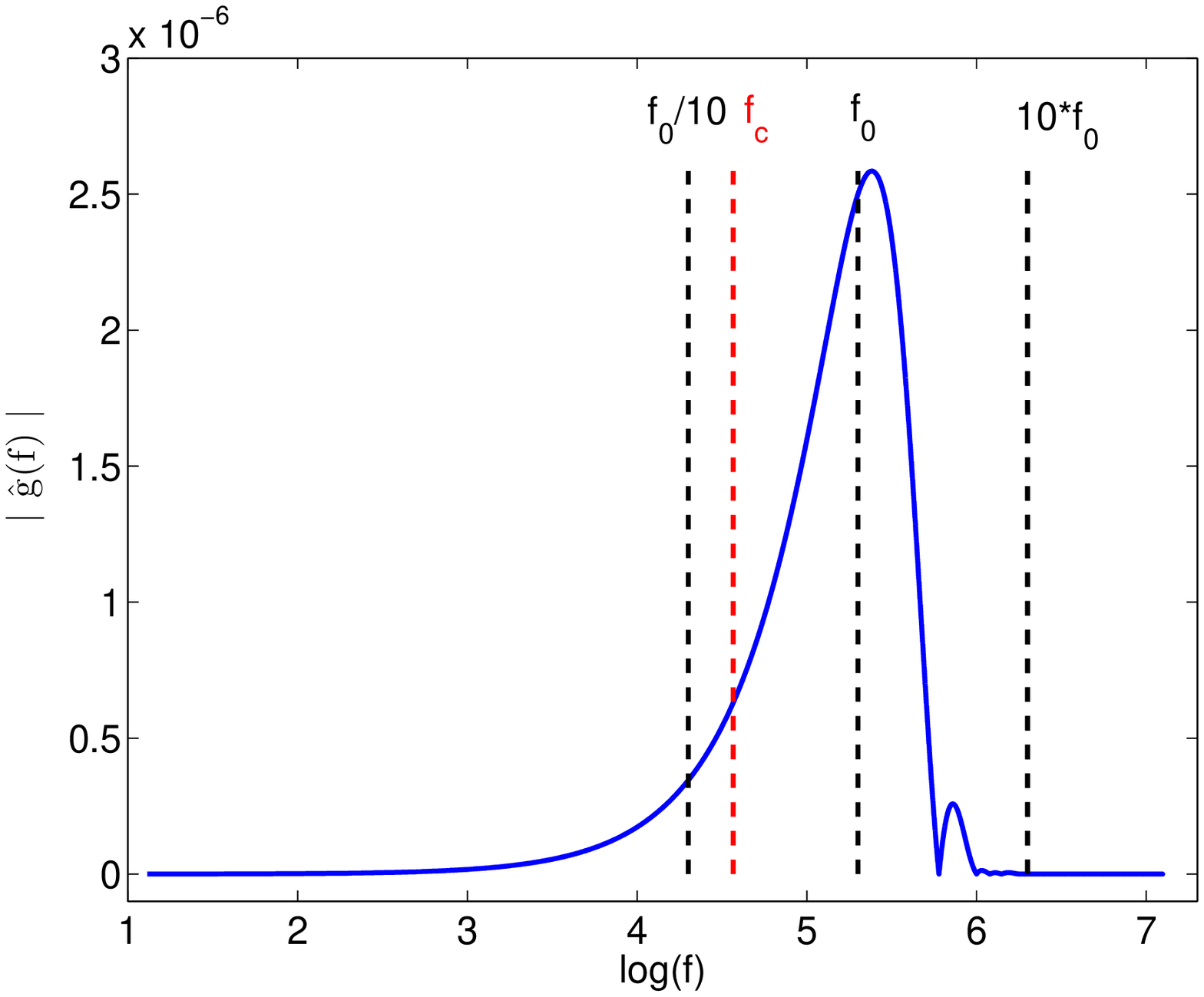}
\end{tabular}
\end{center}
\caption{Time-dependent evolution (left) and spectrum (right) of the source.}
\label{jkps_15khz}
\end{figure}

The physical parameters used in all the numerical experiments, which are given in table \ref{para_phy}, correspond to Berea sandstone saturated with water. Truncated values of the parameters are given: in particular, the viscous characteristic length $\Lambda$ corresponds rigorously to a Pride number $P=0.5$ (\ref{coef_hf}). The unbounded medium is excited by a point source $f_\sigma=g(t)\,h(x)$, with $h(x)=\delta(x)$ in equation (\ref{S1_3}). The time-dependent evolution of the source, $g(t)$ in (\ref{S1_3}), is a $C^6$ combination of truncated sinusoids
\begin{equation}
g(t) = 
\left\lbrace 
\begin{array}{l}
\displaystyle \sin\,(\omega_0t) - \frac{21}{32}\,\sin\,(2\,\omega_0t) + \frac{63}{768}\,\sin\,(4\,\omega_0 t)-\frac{1}{512}\,\sin\,(8\,\omega_0 t) \mbox{ if}\;0\leq t\leq \frac{1}{f_0},\\
\displaystyle 0 \qquad \mbox{otherwise},
\end{array}
\right. 
\label{JKPS_C6}
\end{equation}
with a central frequency $f_0=\frac{\omega_0}{2\,\pi}=200$ kHz. Adopting the high-frequency regime is therefore completely justified since $f_0 \simeq 5 \times f_c$. Figure \ref{jkps_15khz} shows the time-dependent evolution and spectrum of the source.

The computational domain $[-0.04,0.04]$ m is discretized with $N_x$ grid points, and the time step is deduced from (\ref{CFL_dsplit}), taking  $\Upsilon = 0.9$. No special care is applied to simulate outgoing waves (with PML, for instance), since the simulations are stopped before the waves have reached the edges of the computational domain. The numerical experiments are performed on an Intel Core i7 processor at $2.80$ GHz. 

Exact solutions of time-domain Biot-JKD equations have been derived in the literature \cite{FELLAH04}, but not for Biot-DA. Therefore, we compute reference solutions of both Biot-JKD and Biot-DA thanks to standard tools of Fourier analysis: the Green functions of (\ref{S1}) or (\ref{syst2}) are determined in the harmonic regime. Then, the Cauchy residue theorem and numerical inverse Fourier transforms ($N_f = 9.6\,10^{5}$ modes and a frequency step $\Delta\,f \simeq 13$ Hz) yield the semi-analytical solutions. 


\subsection{Test 1: Biot-DA}\label{SecExpTest1}

\begin{figure}[htbp]
\begin{center}
\begin{tabular}{cc}
\includegraphics[scale=0.35]{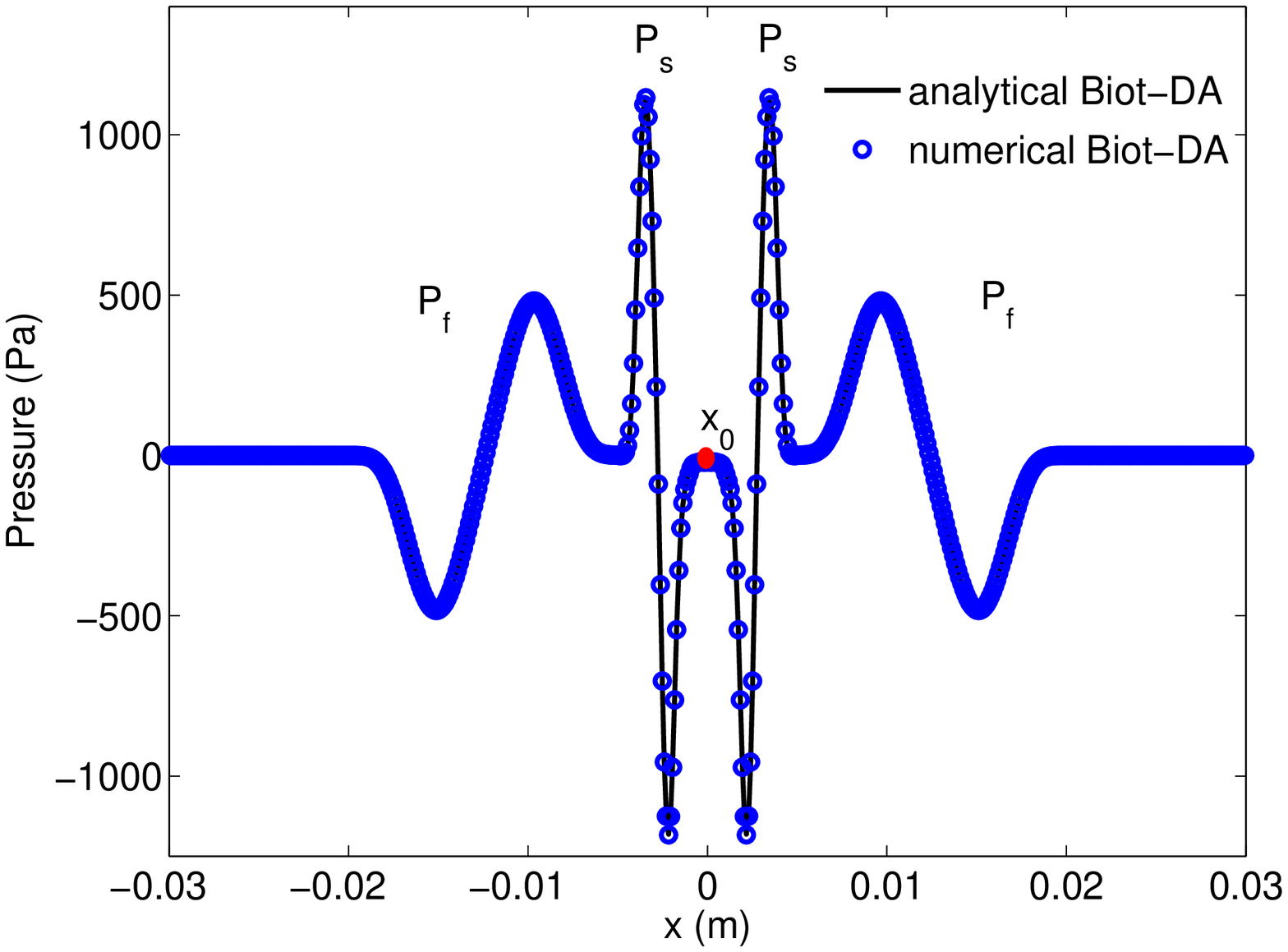} & 
\includegraphics[scale=0.35]{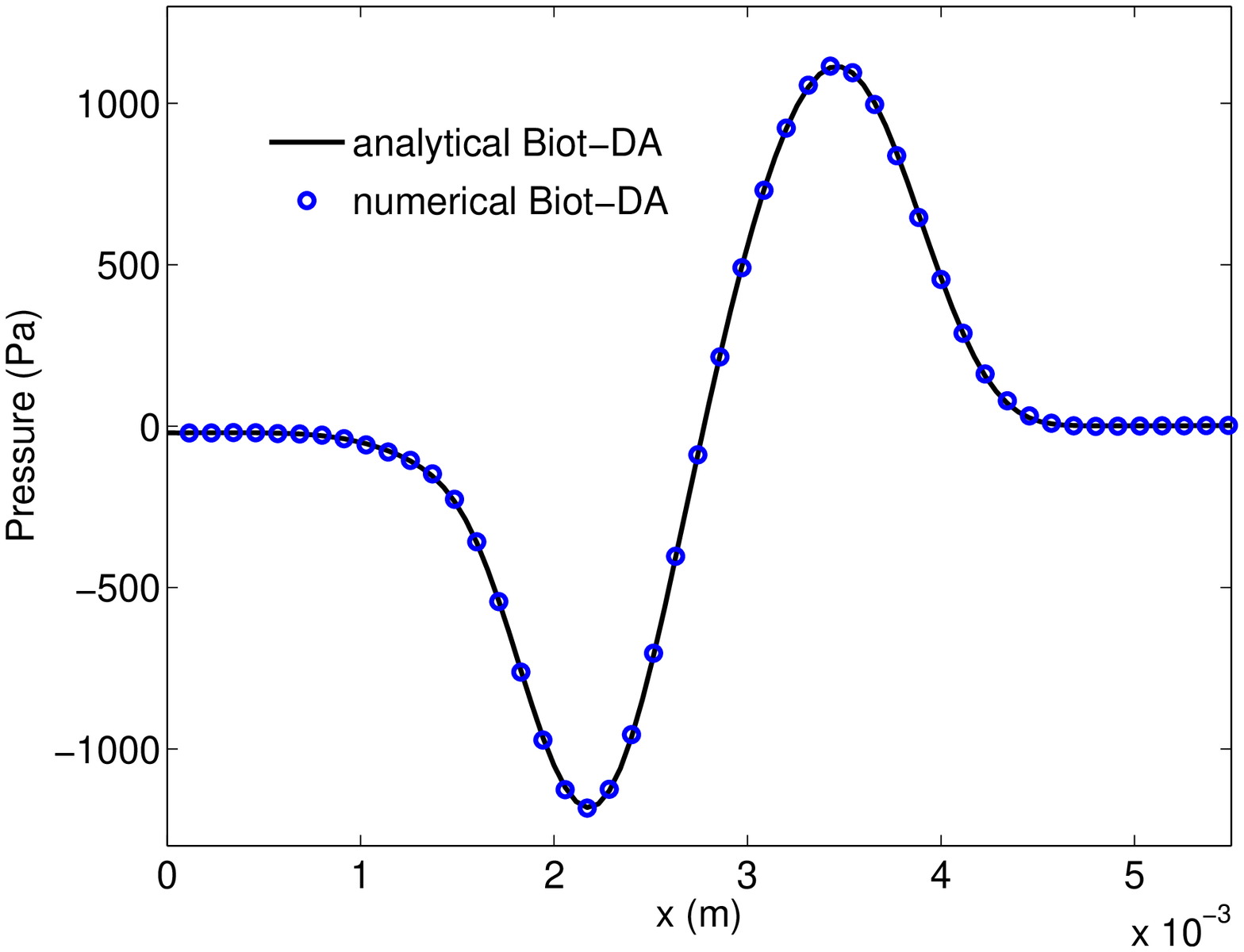}
\end{tabular}
\end{center}
\caption{Test 1. Fast waves $P_f$ and slow waves $P_s$ emitted by a source point at $x_0=0$ m. Comparison between numerical values (circle) and Biot-DA analytical values (solid line) of $p$ at $t_1 \simeq 6.29\,10^{-6}$ s. Right: zoom on the slow wave.}
\label{pression_200khz_AD}
\end{figure}

The aim of this first test is to check the validity of the numerical method presented above using Biot-DA model. The domain is discretized with $N_x=700$ which amounts to 32 points per slow wavelength and 142 points per fast wavelength, and $N=6$ diffusive variables are used. The  source point emits symmetrically rightward and leftward moving fast and slow compressional waves, which are denoted $P_f$ and $P_s$, respectively, in figure  \ref{pression_200khz_AD}. It can be seen from this figure that the numerical and analytical values of the pressure after 200 time steps show excellent agreement.
 
\begin{figure}[htbp]
\begin{center}
\begin{tabular}{cc}
\includegraphics[scale=0.35]{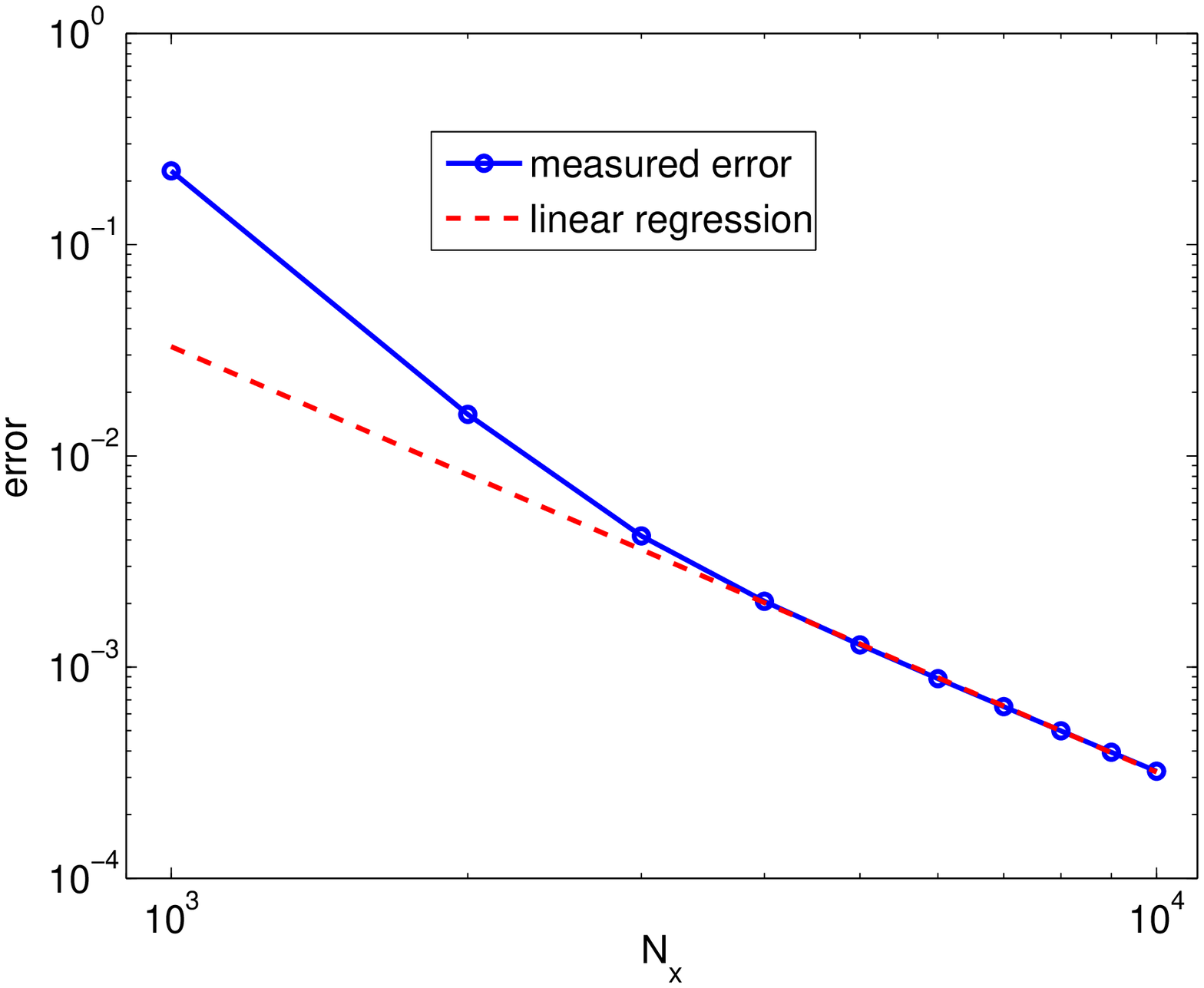}&
\includegraphics[scale=0.35]{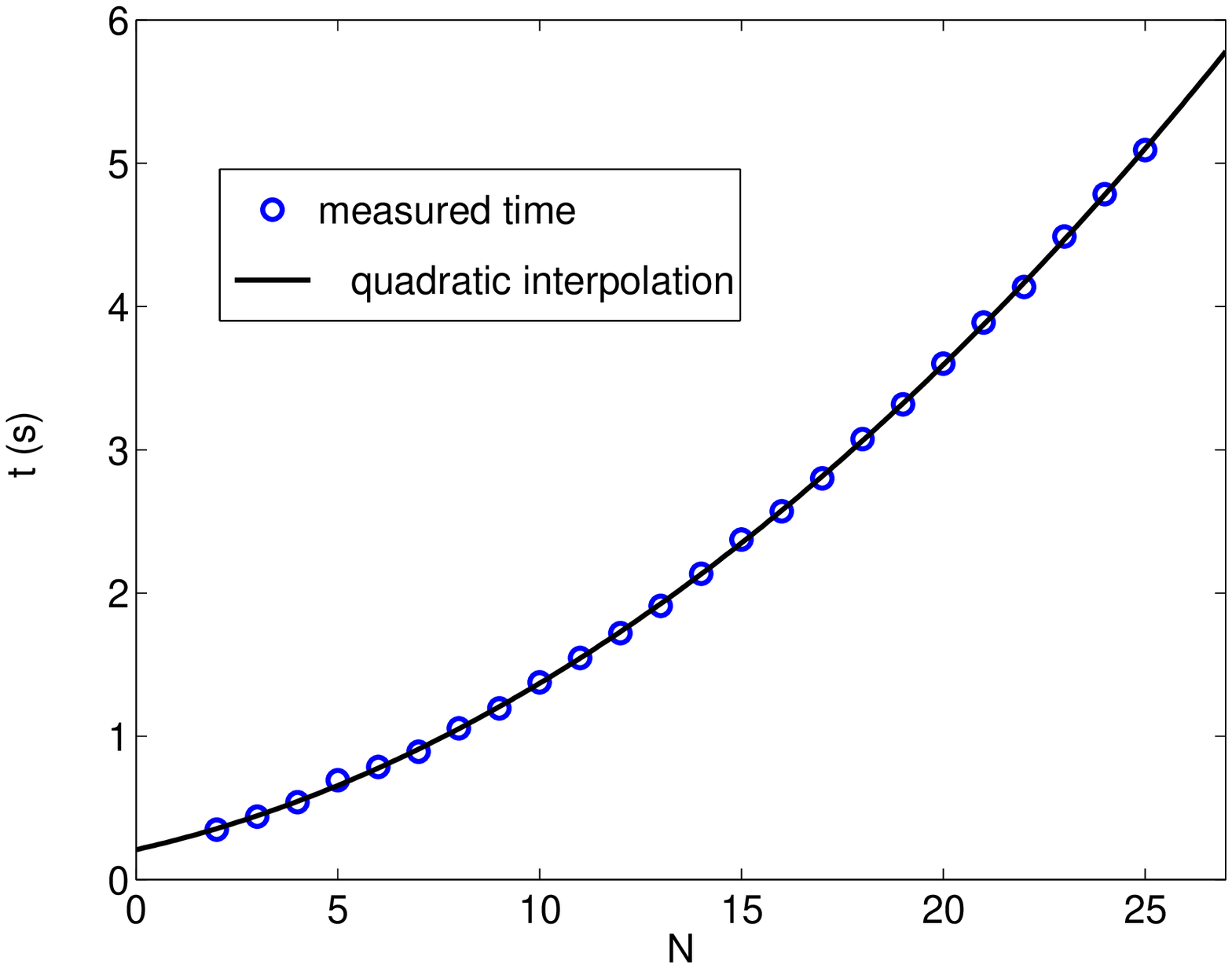}
\end{tabular}
\end{center}
\caption{Test 1: relative error between exact and numerical solutions (left) in terms of the number of grid nodes $N_x$. The dashed line is proportional to $N_x^{-2}$.  CPU time (right) in terms of the number of diffusive variables $N$. }
\label{Fig_ordre_convergence} 
\end{figure}

The error between the exact and numerical solutions  will be measured  in the $L_2$ norm in the domain $[-0.04,0.04]$ m at time $t_1 \simeq 6.29\,10^{-6}$ s. Numerical values of the relative error and convergence order are summed up in table \ref{Table_ordre_convergence} at various values of $N_x$ and given in figure \ref{Fig_ordre_convergence}-(a). The convergence rate obtained by performing a linear regression is $1.97818$, which is very similar to the theoretical second-order of the global algorithm.

\begin{table}[htbp]
\begin{center}
\begin{small}
\begin{tabular}{|r|l|c|}
\hline
\rule[-1mm]{0mm}{5mm} $N_x$ & Error $L_2$ & Order \\ \hline
\rule[-1mm]{0mm}{5mm} 1000  & $1.660\,10^{-1}$ & - \\
\rule[-1mm]{0mm}{3mm} 2000  & $1.554\,10^{-2}$ & $3.417$ \\
\rule[-1mm]{0mm}{3mm} 3000  & $5.939\,10^{-3}$ & $2.372$ \\
\rule[-1mm]{0mm}{3mm} 4000  & $3.300\,10^{-3}$ & $2.043$ \\
\rule[-1mm]{0mm}{3mm} 5000  & $2.121\,10^{-3}$ & $1.981$ \\
\rule[-1mm]{0mm}{3mm} 6000  & $1.482\,10^{-3}$ & $1.968$ \\
\rule[-1mm]{0mm}{3mm} 7000  & $1.095\,10^{-3}$ & $1.963$ \\
\rule[-1mm]{0mm}{3mm} 8000  & $8.428\,10^{-4}$ & $1.958$ \\
\rule[-1mm]{0mm}{3mm} 9000  & $6.699\,10^{-4}$ & $1.950$ \\
\rule[-1mm]{0mm}{3mm} 10000 & $5.462\,10^{-4}$ & $1.937$ \\ \hline
\end{tabular}
\end{small}
\end{center}
\caption{Test 1: error measurements and convergence orders.}
\label{Table_ordre_convergence}
\end{table}

Figure \ref{Fig_ordre_convergence}-(b) shows the computational time in terms of the number of diffusive variables $N$, with $N_x=700$. The complexity of the scheme in term of diffusive variables is found to be in $\mathcal{O}(N^2)$. 

With $N=6$ diffusive variables, the linear optimization procedure described in section \ref{SecDACoeff} yields: $a_1=-371.44$, $a_2=2332.78$, $a_3=-3109.17$, $a_4=4506.03$, $a_5=-4524.14$ and $a_6=7096.95$. Since some of the coefficients are negative, one cannot confirm that $E$ is a decreasing energy in proposition \ref{PropositionNRJ-AD}. To examine this question numerically, the time evolution of $E_3$ in (\ref{Energie-AD}) and $-dE\,/\,dt$ in (\ref{dEdt-AD}) is shown in figure 8, where it can be seen that $E_3>0$, hence $E>0$, and that $dE\,/\,dt<0$. Despite the negativity of some $a_\ell$, figure 8 indicates that $E$ is a decreasing energy and that Biot-DA is a well-posed problem.

\begin{figure}[htbp]
\begin{center}
\begin{tabular}{cc}
\includegraphics[scale=0.35]{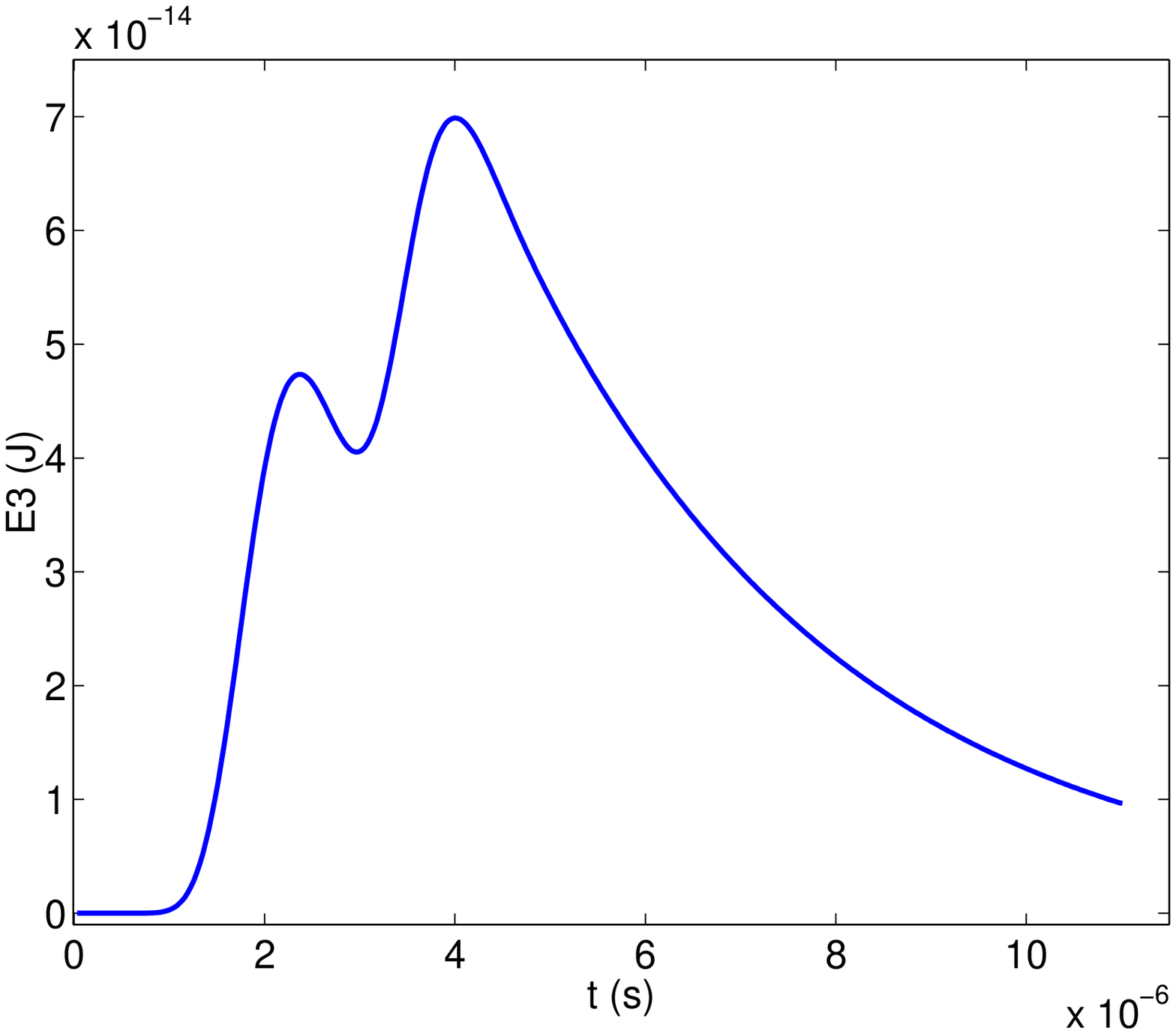} & 
\includegraphics[scale=0.35]{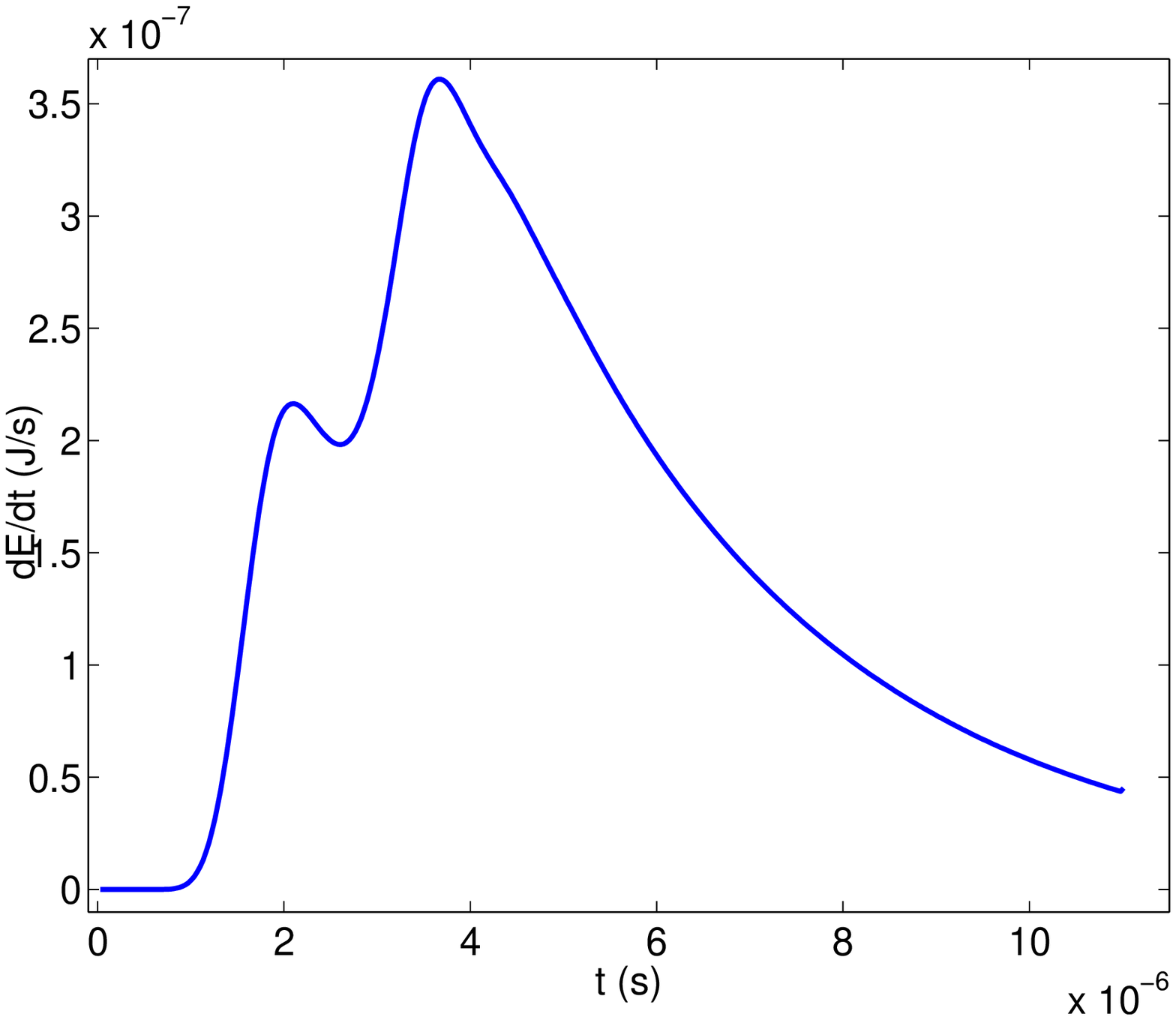}
\end{tabular}
\caption{Test1. Left: time-dependent evolution of $E_3$ (\ref{Energie-AD}); right: time-dependent evolution of $-dE\,/\,dt$ (\ref{dEdt-AD}).}
\end{center}
\label{FigEnergieAD} 
\end{figure}


\subsection{Test 2: Biot-JKD}\label{SecExpTest2}

\begin{figure}[htbp]
\begin{center}
\begin{tabular}{cc}
(a) & (b)\\
\includegraphics[scale=0.35]{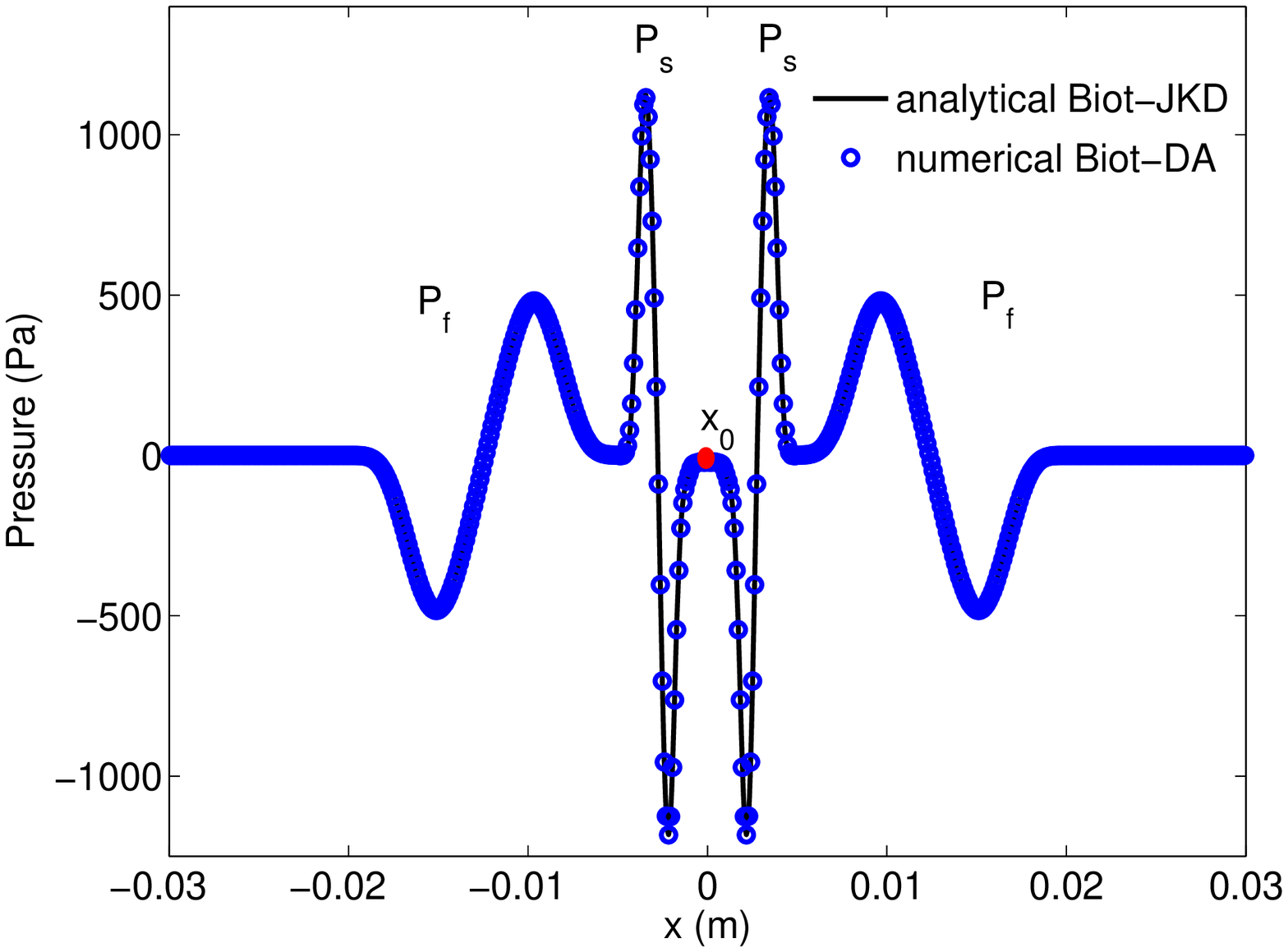} & 
\includegraphics[scale=0.35]{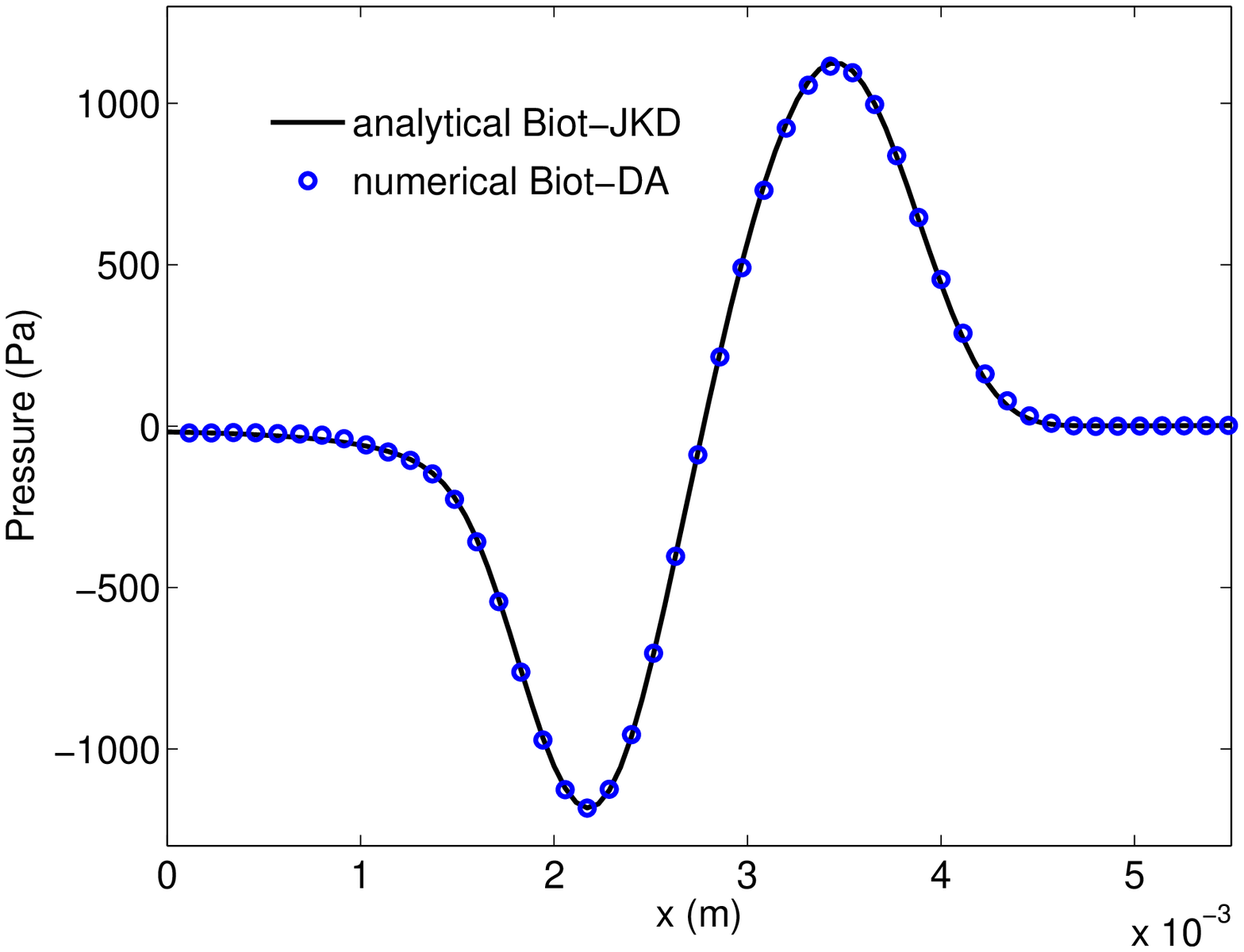}\\
(c) & (d)\\
\includegraphics[scale=0.35]{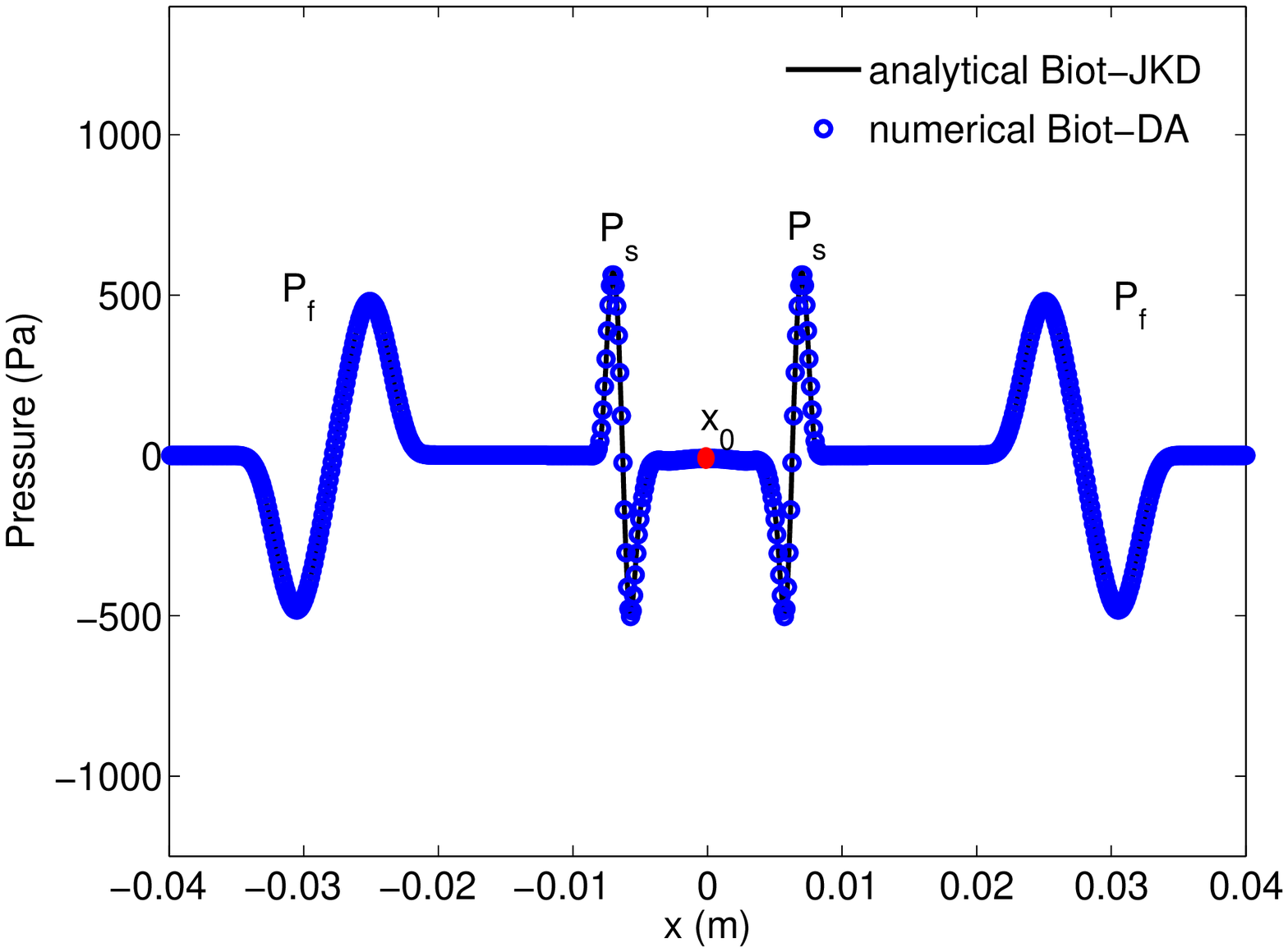} & 
\includegraphics[scale=0.35]{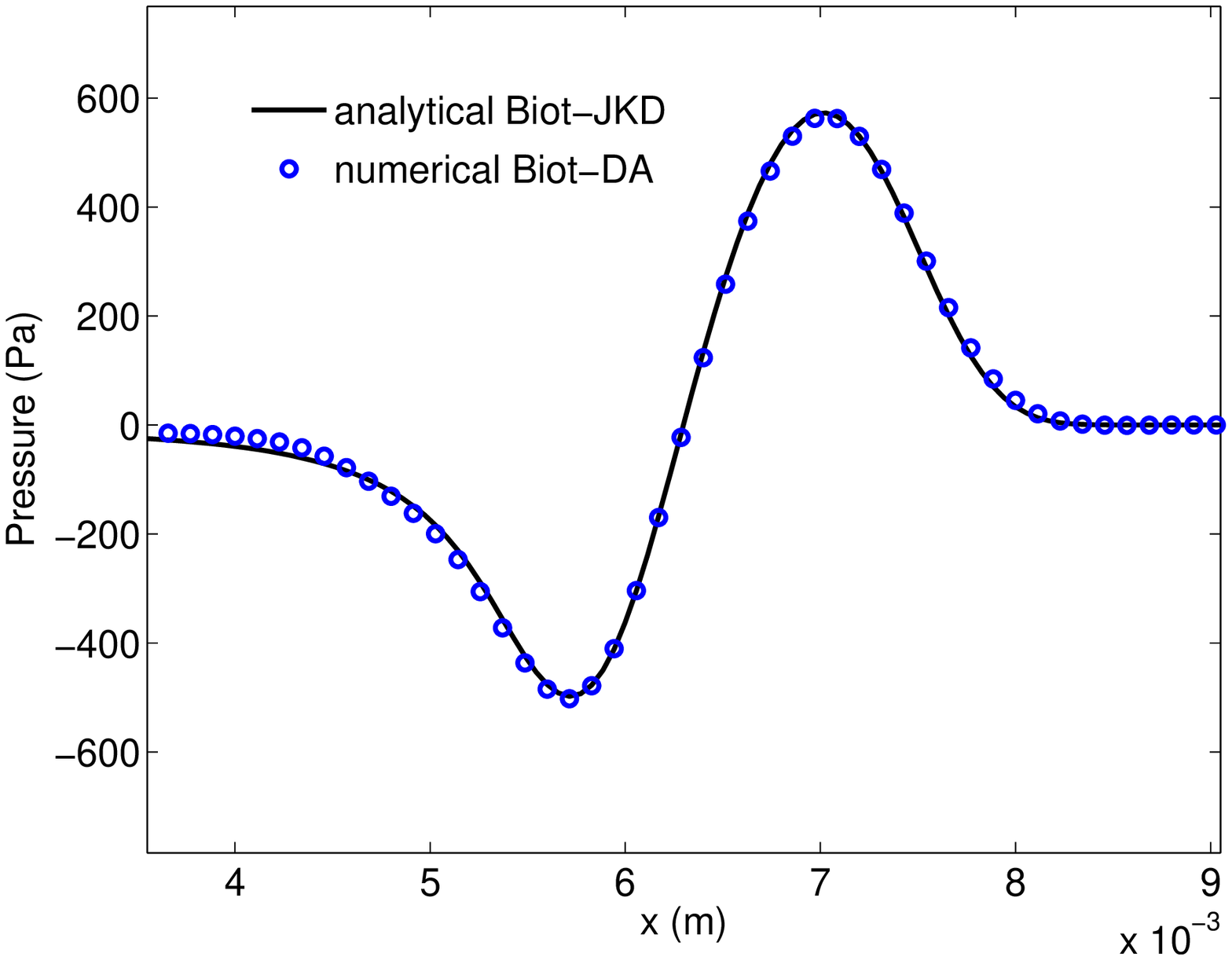}
\end{tabular}
\end{center}
\caption{Test 2. Fast waves $P_f$ and slow waves $P_s$ emitted  by a source point at $t_1 \simeq 6.29\,10^{-6}$ s (a-b) and $t_2 \simeq 1.10\,10^{-5}$ s (c-d). Comparison between the numerical Biot-DA pressure (circle) and the exact Biot-JKD pressure (solid line). Right row: zoom on the slow wave.}
\label{pression_200khz_JKD}
\end{figure}

The aim of the second test is to check the validity of the mathematical and numerical methods used to approximate the physical Biot-JKD model. Figure \ref{pression_200khz_JKD} compares the numerical pressure obtained with the Biot-DA model with the analytical pressure obtained with the Biot-JKD model, at times $t_1$ and $t_2>t_1$. The dispersion and attenuation of the slow wave can be clearly observed. Excellent agreement is found to exist between the two solutions.

Two errors should be mentioned here: the modeling error $\varepsilon_m$, defined as the difference between the Biot-DA and Biot-JKD models; and the numerical error, $\varepsilon_n$, resulting from the numerical discretization of the Biot-DA model. The total error $\varepsilon_t$ obviously satisfies:
\begin{equation}
\varepsilon_t \leq \varepsilon_m+\varepsilon_n.
\label{erreur_totale}
\end{equation}
Based on section \ref{SecDACoeff}, taking $N=6$ yields $\varepsilon_m=5.48$ \%. In test 1, $\varepsilon _n \simeq 1.70$ \% was measured. At $t_1$, the total error $\varepsilon_t=1.95$ \%, which means that the inequality (\ref{erreur_totale}) is satisfied but not optimally: the overall results are more accurate than those predicted on the basis of the bound (\ref{erreur_totale}). The results of this test confirm that the method presented above efficiently approximates the transient waves modeled by the Biot-JKD model.


\subsection{Test 3: variable medium}\label{SecExpTest3}

The aim of the third test is to establish whether the numerical methods presented in this paper can be used to handle more complex media. As an example, we took the porous medium  with the parameters defined in table \ref{para_phy}, except for the ratio $\eta/\kappa$, which varies linearly from $1.5\,10^{4}$ Pa.s.m$^{-2}$ at $x = -0.04$ m to $5\,10^{9}$  Pa.s.m$^{-2}$ at $x = 0.04$ m. These values are purely numerical and are not based on real data. Some changes had to be made to the method in comparison with that used in the homogeneous case:
\begin{itemize}
\item at a given level of accuracy $\varepsilon_m$, the most-penalizing number of diffusive variables $N$ has to be determined;
\item the coefficients $a_{\ell}$ have to be computed and stored at each grid point.
\end{itemize}
In (\ref{syst_hyperbolique}), the diffusive matrix ${\bf S}$ therefore differs between the grid points. In this example, the propagation matrix ${\bf A}$ remains unchanged since only the diffusive part is modified. When dealing with a real continuously variable medium, which occurs in the case of many applications \cite{GROBY11}, the present ADER scheme would also have to be modified in order to handle the spatial changes in the matrix ${\bf A}$ accurately.

\begin{figure}[htbp]
\begin{center}
\begin{tabular}{c}
\includegraphics[scale=0.35]{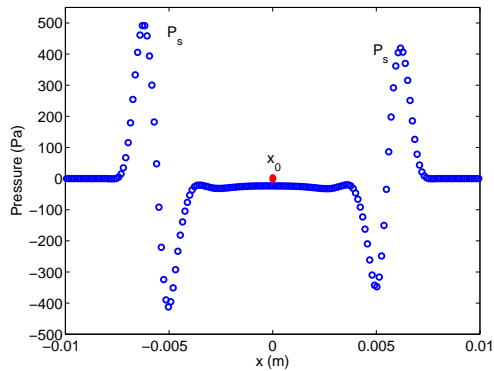}
\end{tabular}
\end{center}
\caption{Test 3. Pressure at $t_1 \simeq 6.29\,10^{-6}$ s (zoom on the slow wave).}
\label{milieu_variable}
\end{figure}

Figure (\ref{milieu_variable}) shows the pressure $p$ at $t_1 \simeq 6.29\,10^{-6}$ s. As was to be expected, the rightward-moving slow wave is more strongly attenuated than the leftward-moving one, because the values of $\eta/\kappa$ are higher in the right part of the domain. The present numerical tool therefore provides useful means for computing solutions of this kind, where no analytical expressions are available.


\subsection{Test 4: a 2-D example}\label{SecExpTest4}

\begin{figure}[htbp]
\begin{center}
\begin{tabular}{cc}
\includegraphics[scale=0.35]{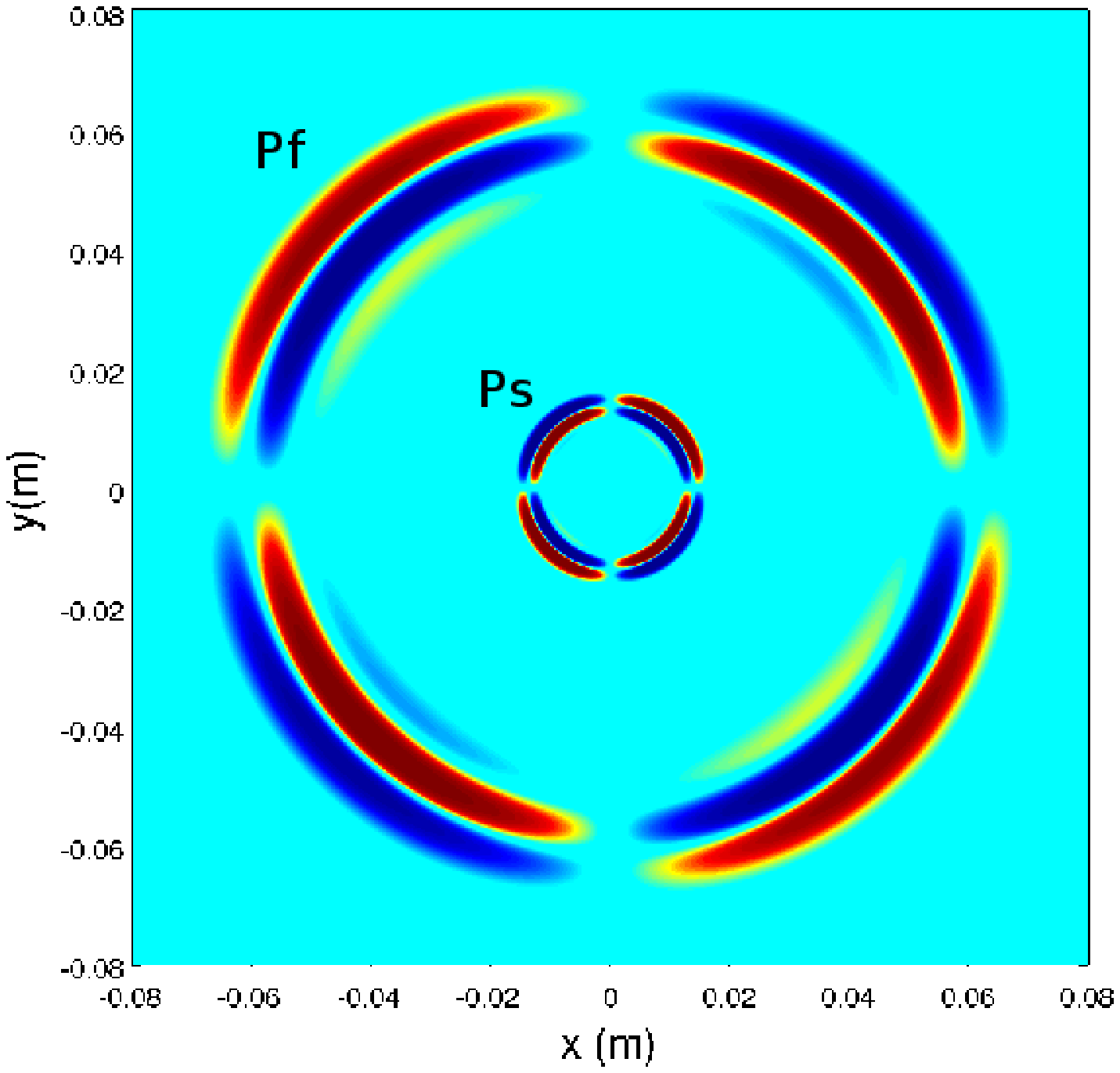} & 
\includegraphics[scale=0.35]{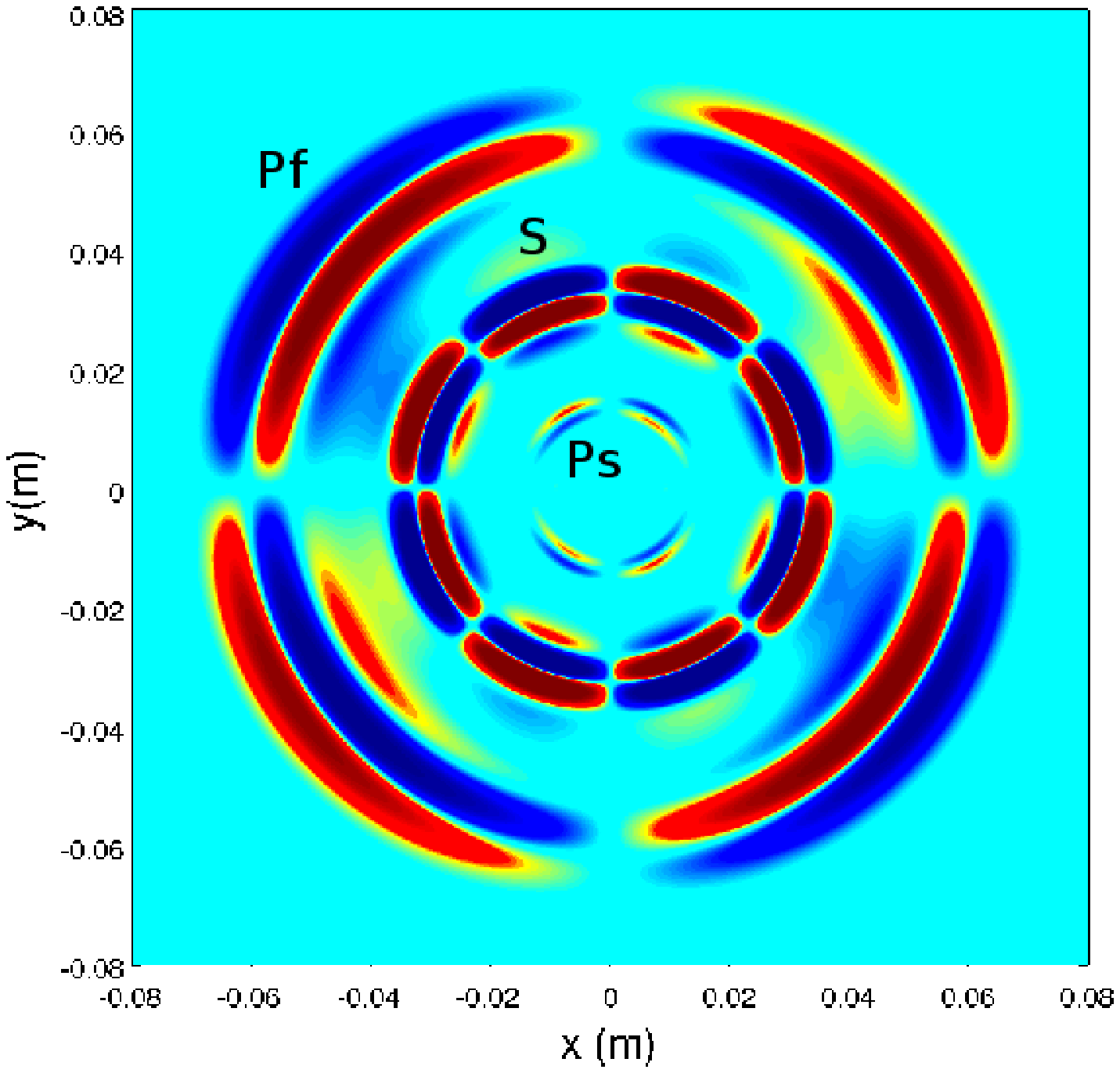}\\
\end{tabular}
\end{center}
\caption{Test 4. Graph of the pressure (left) and the stress component $\sigma_{xx}$ (right) emitted by a source point, at time $t=2.32\,10^{-5}$ s. }
\label{fig2D}
\end{figure}

The one-dimensional method presented here can easily be extended to other dimensions. As a preliminary example, we take a two-dimensional medium with the parameters given in table \ref{para_phy}. The number of physical unknowns increase in this case from 4 to 8, and the equations of motion are also written in the form of a first-order hyperbolic linear system. The propagative part is solved with the ADER 4 numerical scheme. The diffusive part involves an order $1/2$ fractional derivative for each component of the filtration velocity. The computational domain is set at $[-0.08,0.08]^2$ m. A Ricker source point, with a central frequency of $200$ kHz and a time shift $10^{-5}$ s, is localized at point $(0,0)$ and applied to the $\sigma_{xy}$ component of the stress tensor. Applying our method with $N=6$ diffusive variables to a grid of $N_x=N_y=1400$ points gives the results presented in figure \ref{fig2D}. Fast and slow compressional waves are observed as regards the pressure, while the additional shear wave is present in the $\sigma_{xx}$ component of the stress tensor. It is proposed in future studies to address the analytical solution of the 2D Biot-JKD model and to perform an error analysis of the results obtained with the Biot-DA model.

\section{Conclusion}\label{SecConclu}

A numerical method is presented here for simulating transient poroelastic waves in the high-frequency range. The Biot-JKD model, which involves order 1/2 fractional derivatives, was replaced here by an approximate Biot-DA model, which is much more tractable numerically. Contrary to the approach used in \cite{HANYGA05}, the Biot-DA coefficients are determined here using  an optimization procedure, which depends on the frequency range of interest. The number of parameters and the accuracy of the model were quantified. The hyperbolic system of partial differential equations was discretized using efficient tools (Strang splitting and the fourth-order ADER scheme). The stability condition of the numerical scheme is always independent of the parameters involved in the approximate Biot-DA model. Numerical experiments performed in some academic cases (1-D homogeneous media) confirmed the reliability of this approach, and some preliminary simulations (with variable media, or in the 2-D context) show that the method is applicable to complex media.

Some suggestions for future lines of research:
\begin{itemize}
\item \emph{Thermic boundary-layer.} In cases where the saturating fluid is a gas, thermo-mechanical effects have to be taken into account. Extended versions of the Biot-JKD have been developed \cite{LAFARGE97}, involving additional order 1/2 fractional derivatives. The numerical method developed in this paper should lend itself well to working with this model.
\item \emph{Slow shear wave.} A poroelastic theory that accounts properly for the fluid shear stress relaxation has been recently proposed \cite{MULLER11a,MULLER11b}, predicting the existence of a slow shear wave. This additional mode is heavily damped far from the source, but it can play a key role in balance equations near interfaces, as the slow compressional wave. To our knowledge, no time-domain simulations of this model have been proposed so far.
\item \emph{Heterogeneous porous media.} Methods of modeling material interfaces in the context of Cartesian grids have been previously developed, based on an immersed interface method \cite{LOMBARD04}. The possibility of applying this method to porous media in the low frequency range was studied in \cite{CHIAVASSA10,CHIAVASSA11,CHIAVASSA12,MESGOUEZ12}. Work on means of extending this method to the Biot-JKD model is currently in progress.
\end{itemize}


\section*{Acknowledgments}

We are grateful to Denis Matignon (ISAE, Toulouse) for fruitful discussions about fractional derivatives. We also thank Zine Fellah, Erick Ogam, Armand Wirgin (LMA, Marseille), Ga\"elle Lefeuve-Mesgouez and Arnaud Mesgouez (EMMAH, Avignon) for their careful reading of the manuscript.


\appendix

\section{Proof of proposition \ref{PropositionNRJ}}\label{SecProofNRJ-JKD}

The equation (\ref{LCBiot_v}) is multiplied by $v_s$ and integrated 
\begin{equation}
\int_{\mathbb{R}}\left(\rho\,v_s\,\frac{\textstyle \partial\,v_s}{\textstyle \partial\,t}+\rho_f\,v_s\,\frac{\textstyle \partial\,w}{\textstyle \partial\,t}-v_s\,\frac{\textstyle \partial\,\sigma}{\textstyle \partial\,x}\right)\,dx=0.
\label{EProof4}
\end{equation}
The first term in (\ref{EProof4}) is written
\begin{equation}
\int_{\mathbb{R}}\rho\,v_s\,\frac{\textstyle \partial\,v_s}{\textstyle \partial\,t}\,dx=\frac{\textstyle d}{\textstyle dt}\,\frac{\textstyle 1}{\textstyle 2}\int_{\mathbb{R}}\rho\,v_s^2\,dx.
\label{EProof5}
\end{equation}
Integrating by part and using (\ref{LCBiot_p}), we obtain
\begin{equation}
\begin{array}{lll}
\displaystyle
-\int_{\mathbb{R}}v_s\,\frac{\textstyle \partial\,\sigma}{\textstyle \partial\,x}\,dx & = & \displaystyle \int_{\mathbb{R}}\frac{\textstyle \partial\,v_s}{\textstyle \partial\,x}\,\sigma\,dx,\\
[15pt]
& = & \displaystyle \int_{\mathbb{R}}\frac{\textstyle \partial\,\varepsilon}{\textstyle \partial\,t}\,\left(C\,\varepsilon-\beta\,p\right)\,dx,\\
[15pt]
& = & \displaystyle \int_{\mathbb{R}}C\,\varepsilon\frac{\textstyle \partial\,\varepsilon}{\textstyle \partial\,t}\,dx-\int_{\mathbb{R}}\beta\,p\,\frac{\textstyle \partial\,\varepsilon}{\textstyle \partial\,t}\,dx,\\
[15pt]
& = & \displaystyle \frac{\textstyle d}{\textstyle dt}\,\frac{\textstyle 1}{\textstyle 2}\,\int_{\mathbb{R}}C\,\varepsilon^2\,dx-\int_{\mathbb{R}}\beta\,p\,\frac{\textstyle \partial\,\varepsilon}{\textstyle \partial\,t}\,dx,\\
[15pt]
& = & \displaystyle \frac{\textstyle d}{\textstyle dt}\,\left(\frac{\textstyle 1}{\textstyle 2}\,\int_{\mathbb{R}}\frac{\textstyle 1}{\textstyle C}\,\left(\sigma+\beta\,p\right)^2\,dx\right)-\int_{\mathbb{R}}\beta\,p\,\frac{\textstyle \partial\,\varepsilon}{\textstyle \partial\,t}\,dx.
\end{array}
\label{EProof6}
\end{equation}
The equation (\ref{LCBiot_w}) is multiplied by $w$ and integrated
\begin{equation}
\int_{\mathbb{R}}\left(\rho_f\,w\,\frac{\textstyle \partial\,v_s}{\textstyle \partial\,t}+\rho_w\,w\,\frac{\textstyle \partial\,w}{\textstyle \partial\,t}+\frac{\textstyle \eta}{\textstyle \kappa}\,\frac{1}{\sqrt{\Omega}}\,w\,(D+\Omega)^{1/2}w + w\,\frac{\textstyle \partial\,p}{\textstyle \partial\,x}\right)\,dx=0.
\label{EProof7}
\end{equation}
The second term in (\ref{EProof7}) can be written
\begin{equation}
\int_{\mathbb{R}}\rho_w\,w\,\frac{\textstyle \partial\,w}{\textstyle \partial\,t}\,dx=\frac{\textstyle d}{\textstyle dt}\,\frac{\textstyle 1}{\textstyle 2}\,\int_{\mathbb{R}}\rho_w\,w^2\,dx.
\label{EProof7bis}
\end{equation}
Integrating by part and using (\ref{LCBiot_p}), we obtain
\begin{equation}
\begin{array}{lll}
\displaystyle
\int_{\mathbb{R}}w\,\frac{\textstyle \partial\,p}{\textstyle \partial\,x}\,dx & = & \displaystyle -\int_{\mathbb{R}}p\,\frac{\textstyle \partial\,w}{\textstyle \partial\,x}\,dx,\\
[15pt]
& = & \displaystyle \int_{\mathbb{R}}p\,\frac{\textstyle \partial\,\xi}{\textstyle \partial\,t}\,dx,\\
[15pt]
& = & \displaystyle \int_{\mathbb{R}}p\,\frac{\textstyle \partial}{\textstyle \partial\,t}\,\left(\frac{\textstyle 1}{\textstyle m}\,p+\beta\,\varepsilon \right)\,dx,\\
[15pt]
& = & \displaystyle \int_{\mathbb{R}}\frac{\textstyle 1}{\textstyle m}\,p\,\frac{\textstyle \partial\,p}{\textstyle \partial\,t}\,dx+\int_{\mathbb{R}}\beta\,p\,\frac{\textstyle \partial\,\varepsilon}{\textstyle \partial\,t}\,dx,\\
[15pt]
& = & \displaystyle \frac{\textstyle d}{\textstyle dt}\,\left(\frac{\textstyle 1}{\textstyle 2}\,\int_{\mathbb{R}}\frac{\textstyle 1}{\textstyle m}\,p^2\,dx\right)+\int_{\mathbb{R}}\beta\,p\,\frac{\textstyle \partial\,\varepsilon}{\textstyle \partial\,t}\,dx.
\end{array}
\label{EProof8}
\end{equation}
After adding (\ref{EProof4}) and the first term in (\ref{EProof7}), there remains
\begin{equation}
\int_{\mathbb{R}}\rho_f\,\left(v_s\,\frac{\textstyle \partial\,w}{\textstyle \partial\,t} +w\,\frac{\textstyle \partial\,v_s}{\textstyle \partial\,t}\right)\,dx=\frac{\textstyle d}{\textstyle dt}\int_{\mathbb{R}}\rho_f\,v_s\,w\,dx.
\label{EProof9}
\end{equation}
Equations (\ref{EProof4})-(\ref{EProof7}) and the diffusive representation (\ref{derivee_frac}) yield
\begin{equation}
\frac{\textstyle d}{\textstyle dt}\,(E_1 + E_2) = - \int_{\mathbb{R}}\int_{\theta\in\mathbb{R}^+}\frac{\textstyle \eta}{\textstyle \kappa}\,\frac{1}{\pi}\,\frac{1}{\sqrt{\Omega\,\theta}}\,w\,\psi\,d\theta\,dx.
\label{EProof10_bis}
\end{equation}
To calculate the right-hand side of (\ref{EProof10_bis}), equation (\ref{EDO_psi}) is multiplied by $w$ or $\psi$
\begin{subnumcases}{\label{EProof0}}
\displaystyle w\,\frac{\partial \psi}{\partial t} - w\,\frac{\partial w}{\partial t} + (\theta+\Omega)\,w\,\psi-\Omega\,w^2 = 0\, ,\label{EProof0_1}\\
[13pt]
\displaystyle \psi\,\frac{\partial \psi}{\partial t} - \psi\,\frac{\partial w}{\partial t} + (\theta+\Omega)\,\psi^2-\Omega\,w\,\psi = 0 .\label{EProof0_2}
\end{subnumcases}
After performing some algebraic operations on (\ref{EProof0_2}), (\ref{EProof0_1}) and (\ref{EProof10_bis}), one easily obtains the relation  (\ref{EdEdt}). It remains to prove that $E$ is a positive definite quadratic form. This is obviously so for $E_2$ and $E_3$. Concerning $E_1$, we write
\begin{equation}
\begin{array}{lll}
\displaystyle \Delta &=& \displaystyle \frac{\textstyle 1}{\textstyle 2}\,\rho\,v_s^2+\frac{\textstyle 1}{\textstyle 2}\,\rho_w\,w^2+\rho_f\,v_s\,w,\\
[12pt]
& = & \displaystyle \frac{1}{2}\,{\bf X}^T\,{\bf H}\,{\bf X},
\end{array}
\label{EProof18}
\end{equation}
where
\begin{equation}
{\bf X} = 
\left( 
\begin{array}{l} 
v_s \\ 
w 
\end{array} 
\right)\hspace{1cm} 
{\bf H}=
\left( 
\begin{array}{ll} 
\rho & \rho_f \\ 
\rho_f & \rho_w 
\end{array} 
\right).
\end{equation}
Taking $\mathcal{S}$ and $\mathcal{P}$  to denote the sum and the product of the eigenvalues of  matrix ${\bf H}$, we obtain 
\begin{equation}
\begin{array}{l}
\mathcal{P} = \det \,{\bf H} = \rho\,\rho_w - \rho_f^2 = \chi \; >\;0,\\
[10pt]
\mathcal{S} = \mbox{Tr}\,{\bf H} = \rho + \rho_w \; >\;0.
\end{array}
\end{equation}
The two eigenvalues of ${\bf H}$ are therefore positive, which proves that $\Delta$ is definite positive and completes the proof.


\section{Proof of proposition \ref{PropositionStabExp}}\label{SecProofStab}

From (\ref{syst2}), the system of diffusive evolution equations writes
\begin{subnumcases}{\label{nrj_split}}
\displaystyle \frac{\partial v_s}{\partial t} = \frac{\rho _f}{\rho}\,\gamma\,\sum \limits _{\ell =1}^{N}a_{\ell } \,\psi _{\ell }\mbox{,}\label{nrj_spli_v}\\
\displaystyle \frac{\partial w}{\partial t} = -\gamma\,\sum \limits _{\ell =1}^{N}a_{\ell }\, \psi _{\ell }\mbox{,}\label{nrj_spli_w}\\
\displaystyle \frac{\partial \sigma}{\partial t} = 0\mbox{,}\label{nrj_spli_s}\\
[5pt]
\displaystyle \frac{\partial p}{\partial t} = 0\mbox{,}\label{nrj_spli_p}\\
\displaystyle \frac{\partial \psi _j}{\partial t} = \Omega \,w - \gamma\,\sum \limits _{\ell =1}^{N}a_{\ell}\, \psi _{\ell} -(\theta_j + \Omega ) \,\psi _j \mbox{,}\qquad j = 1,...,N\mbox{.}\label{nrj_spli_psi}
\end{subnumcases}
Equation (\ref{nrj_spli_w}) is multiplied by $w$ and (\ref{nrj_spli_psi}) is multiplied by $\psi_j$
\begin{subnumcases}
\displaystyle w \,\frac{\partial\,w}{\partial\,t} = -\gamma\,w\,\sum\limits _{\ell =1}^N a_{\ell}\,\psi _{\ell},\label{nrj_split_proof1}\\
\displaystyle \psi_j\,\frac{\partial\,\psi_j}{\partial\,t} = \Omega\,w\,\psi_j - \gamma\,\psi_j\,\sum\limits _{\ell =1}^N a_{\ell}\,\psi _{\ell} - (\theta_j+\Omega)\,\psi_j^2,\quad j = 1,...,N\label{nrj_split_proof2}.
\end{subnumcases}
Summing (\ref{nrj_split_proof1}) and (\ref{nrj_split_proof2}) gives
\begin{equation}
w\,\frac{\partial\,w}{\partial\,t} + \psi_j\,\frac{\partial\,\psi_j}{\partial\,t} = \Omega\,w\,\psi_j - (\theta_j+\Omega)\,\psi_j^2 -\gamma\,(w+\psi_j)\,\sum\limits _{\ell =1}^N a_{\ell}\,\psi _{\ell},\qquad j = 1,...,N.
\label{nrj_split_proof3}
\end{equation}
The left-hand-side of (\ref{nrj_split_proof3}) is equal to $\frac{d}{dt}\,\frac{1}{2}\,(w^2+\psi_j^2)$. Then (\ref{nrj_spli_w}) is multiplied by $\psi_j$ and (\ref{nrj_spli_psi}) is multiplied by $w$ 
\begin{subnumcases}
\displaystyle \psi _j\,\frac{\partial\,w}{\partial\,t} = -\gamma\,\psi_j\,\sum\limits _{\ell =1}^N a_{\ell}\,\psi _{\ell},\qquad j = 1,...,N,\label{nrj_split_proof4}\\
\displaystyle w\,\frac{\partial\,\psi_j}{\partial\,t} = \Omega\,w^2 - \gamma\,w\,\sum\limits _{\ell =1}^N a_{\ell}\,\psi _{\ell} -(\theta_j+\Omega)\,w\,\psi_j,\qquad j = 1,...,N.\label{nrj_split_proof5}
\end{subnumcases}
Summing (\ref{nrj_split_proof4}) and (\ref{nrj_split_proof5}) gives
\begin{equation}
w\,\frac{\partial\,\psi_j}{\partial\,t} + \psi_j\,\frac{\partial\,w}{\partial\,t} = \Omega\,w^2 - (\theta_j+\Omega)\,w\,\psi_j - \gamma\,(w+\psi_j)\,\sum\limits _{\ell =1}^N a_{\ell}\,\psi _{\ell},\qquad j = 1,...,N.
\label{nrj_split_proof6}
\end{equation}
The left-hand-side of (\ref{nrj_split_proof6}) writes $\frac{\partial}{\partial\,t}\,(w\,\psi_j)$. Elementary calculations on (\ref{nrj_split_proof6}) and (\ref{nrj_split_proof3}) yield ($j = 1,...,N$)
\begin{equation}
\frac{d}{dt}\left( \frac{1}{2}\,(w^2 + \psi_j^2)-w\,\psi_j\right)  = -\left( \Omega\,w^2 - (\theta_j+2\,\Omega)\,w\,\psi_j + (\theta_j+\Omega)\,\psi_j^2\right).
\label{nrj_split_proof7}
\end{equation}
Taking
\begin{equation}
\left\lbrace 
\begin{array}{ll}
\displaystyle E_j & \displaystyle = \frac{1}{2}\,(w - \psi_j)^2 >0,\\
[10pt]
\displaystyle E & \displaystyle = \sum\limits _{j =1}^N E_j >0,\\
[10pt]
\displaystyle {\bf X}_j & \displaystyle =\left( \begin{array}{c}w\\\psi_j\end{array}\right) \\
[13pt]
\displaystyle {\bf H}_j & \displaystyle = \left( \begin{array}{cc}
\Omega & -(\theta_j+2\,\Omega) \\
[5pt]
0 & \theta_j+\Omega,
\end{array}\right) ,
\end{array}
\right. 
\label{nrj_split_proof8}
\end{equation}
and summing the relations (\ref{nrj_split_proof7}) for $j=1,\cdots,N$ yields
\begin{equation}
\frac{dE}{dt} = -\sum\limits _{j =1}^N {\bf X}_j^T\,{\bf H}_j\,{\bf X}_j.
\label{nrj_split_proof9}
\end{equation} 
Since the  matrix ${\bf H}_j$ is triangular, its two eigenvalues are $\Omega >0$ and $\theta_j+\Omega >0$. The quadratic form ${\bf X}_j^T\,{\bf H}_j\,{\bf X}_j$ is therefore definite and positive, which means that the left-hand-side of (\ref{nrj_split_proof9}) is strictly negative. The energy $E$ derived from system (\ref{nrj_split}) is therefore decreasing, and hence the system (\ref{nrj_split}) is well-posed.


\end{document}